\documentclass[journal, hidelinks]{IEEEtran}
\usepackage{caption}
\usepackage{algorithm}
\usepackage{algorithmicx}
\usepackage{algpseudocode}
\usepackage{array}
\usepackage{textcomp}
\usepackage{stfloats}
\usepackage{url}
\usepackage{verbatim}
\usepackage{graphicx}
\usepackage{cite}
\usepackage{orcidlink}
\usepackage{amsmath,amssymb,mathtools}
\usepackage[english]{babel}
\usepackage{bm}
\usepackage{csquotes}
\usepackage{lipsum}
\usepackage{subfigure}   
\usepackage{float}       
\usepackage{graphicx}   
\usepackage[acronym,shortcuts]{glossaries}
\usepackage{caption}
\usepackage{booktabs}
\usepackage{microtype}
\usepackage{balance}

\newcommand{\trans}[0]{^{\mathsf{T}}}
\newcommand{\herm}[0]{^{\mathrm{H}}}

\newcommand{\lm}[0]{\lambda}

\newacronym{MIMO}{MIMO}{multiple-input multiple-output}
\newacronym{MU-mMIMO}{MU-mMIMO}{multi-user massive MIMO}
\newacronym{CF-mMIMO}{CF-mMIMO}{cell-free massive multiple-input multiple-output}
\newacronym{SotA}{SotA}{state-of-the-art}
\newacronym{RF}{RF}{radio frequency}
\newacronym{SNR}{SNR}{signal-to-noise ratio}
\newacronym{SINR}{SINR}{signal-to-interference-plus-noise ratio}
\newacronym{iid}{i.i.d}{independent and identically distributed}
\newacronym{AP}{AP}{access point}
\newacronym{UE}{UE}{user equipment}
\newacronym{CSI}{CSI}{channel state information}
\newacronym{TDD}{TDD}{time-division duplex}
\newacronym{MMSE}{MMSE}{minimum mean square error}
\newacronym{MF}{MF}{matched filter}
\newacronym{MSE}{MSE}{mean squared error}
\newacronym{MRT}{MRT}{maximum ratio transmission}
\newacronym{CPU}{CPU}{central processing unit}
\newacronym{OB}{OB}{oblique}
\newacronym{MUI}{MUI}{multiple access interference}
\newacronym{DS}{DS}{desired signal}
\newacronym{BU}{BU}{beamforming uncertainty}
\newacronym{NP}{NP}{nondeterministic polynomial time}
\newacronym{RPE}{RPE}{radar parameter estimation}
\newacronym{ISAC}{ISAC}{integrated sensing and communication}
\newacronym{BER}{BER}{bit error rate}
\newacronym{LMMSE}{LMMSE}{linear minimum mean square error}
\newacronym{GaBP}{GaBP}{Gaussian belief propagation}
\newacronym{EP}{EP}{expectation propagation}
\newacronym{IO}{I/O}{input/output}
\newacronym{sIC}{sIC}{soft interference cancellation}
\newacronym{SGA}{SGA}{standard Gaussian approximation}
\newacronym{PDF}{PDF}{probability density function}
\newacronym{QPSK}{QPSK}{quadrature phase shift keying}
\newacronym{wlg}{wlg}{without loss of generality}
\newacronym{FP}{FP}{fractional programming}
\newacronym{BS}{BS}{base station}
\newacronym{DAC}{DAC}{digital-to-analog converter}
\newacronym{ADC}{ADC}{analog-to-digital converter}
\newacronym{WF}{WF}{Wiener filter}
\newacronym{ZF}{ZF}{zero forcing}
\newacronym{QCQP}{QCQP}{quadratically constrained quadratic programming}
\newacronym{AMP}{AMP}{approximate message passing}
\newacronym{IDLS}{IDLS}{iterative discrete least squares}
\newacronym{BP}{BP}{belief propagation}
\newacronym{CDM}{CDM}{coordinate descent method}
\newacronym{QP}{QP}{quantized precoding}
\newacronym{EVM}{EVM}{error vector magnitude}
\newacronym{MM}{MM}{majorization-minimization}
\newacronym{MQP}{MQP}{multibit quantized precoding}
\newacronym{GNC}{GNC}{graduated non-convexity}
\newacronym{MP}{MP}{message-passing}

\begin{document}
\title{Multibit Quantized Precoding for MU-mMIMO}

\author{Getuar Rexhepi\textsuperscript{\orcidlink{0009-0002-3268-522X}},~\IEEEmembership{Graduate Student Member,~IEEE}, Shreesal Shrestha \textsuperscript{\orcidlink{0009-0006-5320-7234}},~\IEEEmembership{Student Member,~IEEE},\\  Christoph Studer\textsuperscript{\orcidlink{0000-0001-8950-6267}},~\IEEEmembership{Senior Member,~IEEE}, and Giuseppe Thadeu Freitas de Abreu\textsuperscript{\orcidlink{0000-0002-5018-8174}}~\IEEEmembership{Senior Member,~IEEE\vspace{-3.5ex}}
\thanks{Getuar Rexhepi, Shreesal Shrestha, and Giuseppe Thadeu Freitas de Abreu are with the School of Computer Science and Engineering, Constructor University (previously Jacobs University Bremen), Campus Ring 1, 28759 Bremen, Germany. Emails: [grexhepi, shrshrestha, gabreu]@constructor.university.}

\thanks{Christoph Studer is with the Department of Information Technology and Electrical Engineering, ETH Zürich, 8092 Zürich, Switzerland. Email: cstuder@ethz.ch.}}

\maketitle

\begin{abstract}
We propose a novel
multibit quantized precoding method for the downlink of multi-user massive MIMO systems with low-resolution digital-to-analog converters.
{The new method, termed \ac{MQP}, enforces the finite-alphabet constraint through an $\ell_0$-norm penalty, approximated by a smooth surrogate so as to yield a reformulated problem, which is then convexized via fractional programming, ultimately extending quantized precoding beyond 1-bit alphabets.}
{The regularization parameter of the proposed method is selected via a discrepancy principle integrated with graduated non-convexity continuation, resulting in a principled and reproducible hyperparameter tuning method and an efficient iterative algorithm with a closed-form, least-squares-type update per iteration.
In order to further reduce the computational complexity of the method, we include a  \ac{GaBP} step for turning the least-squares update in linear-time.}
{Simulations performed for systems with different sizes demonstrate that both methods, namely the \ac{MQP} with and without \ac{GaBP}, achieve competitive or superior error-rate performance compared to state-of-the-art quantized precoding algorithms under various channel conditions.}
\end{abstract}

\begin{IEEEkeywords}
Massive MIMO, quantized precoding, low-resolution DACs, one-bit quantization, fractional programming
\end{IEEEkeywords}

\glsresetall

\vspace{-4ex}
\section{Introduction}
\label{sec:intro}

The rapid growth of connected devices and bandwidth-hungry applications has driven an unprecedented surge in mobile data traffic, placing increasingly stringent demands on the capacity and energy efficiency of wireless networks~\cite{SaadIEEENetwork2020, GiordaniComMag2020, WangComST2023, Ericsson2024}.
In order to meet these requirements, fifth-generation (5G) and beyond wireless systems have embraced \ac{MU-mMIMO} as a key enabling technology~\cite{Larsson2014}.
Since the seminal contribution of~\cite{Marzetta2010}, \ac{MU-mMIMO} has reached widespread recognition  in wireless research, owing to its remarkable potential for jointly improving spectral and energy efficiency.
By equipping \acp{BS} with hundreds of antenna elements, \ac{MU-mMIMO} systems is able to concurrently serve many users over shared time-frequency resources, exploiting the spatial degrees of freedom available in the propagation environment to realize substantial throughput improvements \cite{Bjornson2014, Bjornson2017}.
Nevertheless, deploying \ac{MU-mMIMO} in practice introduces hardware design challenges, most notably the power consumption and cost associated with high-resolution \acp{DAC} at the \ac{BS} \cite{gustavsson2014}.

In conventional massive \ac{MIMO} architectures, each antenna element is driven by a dedicated \ac{RF} chain incorporating a high-resolution \ac{DAC} that converts the digitally precoded baseband signal into an analog transmit waveform.
As the number of antenna elements increases, the total power drawn by these \acp{DAC} becomes a dominant and often unsustainable fraction of the total \ac{BS} power budget.
This burden is further aggravated by the well-known exponential scaling of \ac{DAC} power dissipation with resolution~\cite{adc_survey}, rendering high-fidelity solutions increasingly impractical in large-scale deployments and motivating a paradigm shift toward coarsely quantized transmitter hardware.

\vspace{-2ex}
\subsection{Related Prior Art}

Low-resolution \acp{DAC} offer compelling advantages for \ac{MU-mMIMO} base stations, including substantially reduced power consumption, simplified circuit design, and lower component cost.
In the extreme case of 1-bit \acp{DAC}, which produce only two output levels, dramatic reductions in hardware complexity and power dissipation are achievable, while simultaneously enabling highly efficient all-digital beamforming architectures.
However, classical precoders such as \ac{MRT} and \ac{ZF}, derived under the assumption of continuous-valued transmit signals, suffer substantial performance degradation when subjected to coarse quantization, since the resulting quantization noise introduces inter-user interference.
In this context, the design of nonlinear precoding algorithms that explicitly account for the discrete nature of the transmit alphabet has emerged as a critical research direction \cite{Jacobsson2017quantized, Castaneda2017, LiTWC2018}, with the goal of recovering much of the performance lost to quantization while retaining the hardware and energy efficiency benefits of low-resolution \acp{DAC}.
Nevertheless, the combinatorial nature of the discrete precoding problem renders it NP-hard in general, motivating a rich body of work on tractable approximations and relaxations that balance performance and complexity.

{However, the widely adopted surrogate formulation for 1-bit precoding, which minimizes a regularized \ac{MSE} objective \cite{Jacobsson2017quantized, Castaneda2017, LiTWC2018}, replaces the intractable combinatorial search over discrete transmit alphabets with a tractable continuous relaxation.}
{While this approach yields computationally efficient solutions that approximately account for quantization effects, it suffers from two fundamental shortcomings.}
{First, the regularization weight in the problem formulation lacks theoretical grounding and must be selected heuristically, typically via cross-validation.}
As a result, significant performance degradation can occur under varying channel conditions, system configurations, or operating \acp{SNR}.
{Second, the framework does not generalize systematically to
higher-order \ac{DAC} resolutions, which are crucial for systems to meet the aforementioned stringent rate and energy-efficiency requirements \cite{SaadIEEENetwork2020, GiordaniComMag2020, WangComST2023,Ericsson2024}.}

In order to address the combinatorial nature of finite-alphabet precoding more directly, \ac{CDM}, originally developed in the context of channel equalization, can be adapted for precoding design \cite{Chen2022OneBitMMSE}, since the precoding problem can be viewed as a dual of the equalization problem.
In such a framework, the precoding vector is optimized one element at a time, cycling through the antenna indices and selecting, for each element, the alphabet symbol that minimizes a carefully chosen objective while holding all remaining elements fixed.
As a result, \ac{CDM} benefits from a low per-iteration cost, natural compatibility with arbitrary discrete alphabets, and monotone convergence guarantees.
Nonetheless, to achieve competitive performance, \ac{CDM} requires initialization with a high-quality starting point, typically obtained from a linear precoder such as \ac{WF} or \ac{ZF}, {which requires matrix inversions of complexity scaling
rapidly with the number of antennas}~\cite{Saxena2017}.
{Furthermore, \ac{CDM} is not robust to channel estimation errors, making it less suitable for practical deployments, where perfect \ac{CSI} is unavailable.}

\vspace{-3ex}
\subsection{Proposed Approach and Contributions}

{Motivated by the above observations, we close the gap between the non-generalizable surrogate formulation and the high-complexity \ac{CDM} approach, by proposing \ac{MQP}, a novel discrete-aware precoding framework that explicitly incorporates the finite-alphabet constraint into the optimization objective and can be solved efficiently using an \ac{MP} algorithm.}
To this end, we employ \ac{FP}~\cite{Shen2018}, a powerful majorization technique that transforms a non-convex fractional objective into a sequence of convex subproblems, {each of which can be solved in closed form.}
The technique has proven highly effective in tackling challenging {non-convex} optimization problems across a wide range of wireless communication applications, including massive NOMA detection, reciprocal BD-RIS scattering matrix optimization, and secure ISAC beamforming optimization \cite{Iimori2021Robust, Fidanovski2025BDRIS_FP, Boroujeni2025SecureISAC_FP}.
Combining \ac{FP} with a smooth approximation of the $\ell_0$-norm penalty that enforces the discrete-alphabet constraint, we obtain an efficient iterative algorithm which, similarly to \ac{CDM}, accommodates arbitrary discrete alphabets.

Unlike \ac{CDM}, however, the proposed the \ac{GaBP} variation of \ac{MQP} does not require a good initialization.
In addition, the method operates via \ac{MMSE}-type closed-form updates per iteration, which is more efficient than the search over the alphabet required by \ac{CDM}.
To deal with the high dimensionality of the problem, we leverage a \ac{MP} approach that exploits the factor{-}graph structure of the problem to reduce the per-iteration complexity from cubic to linear in the number of antennas, and the initialization from a matrix inversion to a simple matched filter, or just the zero vector.
Message-passing algorithms have emerged as powerful tools for efficiently solving large-scale {non-convex} inference and optimization problems by exploiting the underlying graph structure.
Their effectiveness has been demonstrated in diverse wireless applications, such as scalable integrated communication and computing receivers and Gaussian belief propagation-based localization \cite{Ranasinghe2024ICCReceiver, Fuhrling2024GBPLocalization}.

In summary, our main contributions are as follows:
\begin{enumerate}
\item We propose \ac{MQP}, a
multibit nonlinear precoding framework that explicitly incorporates the finite-alphabet constraint into the optimization objective and, unlike prior surrogate formulations, applies to \acp{DAC} of arbitrary resolution. {The resulting problem is solved efficiently via a generalized \ac{MMSE} update derived via \ac{FP}, which admits a closed-form least-squares-type solution per iteration.}
\item {We derive a message-passing variant of \ac{MQP} that exploits the factor-graph structure of the problem via \ac{GaBP}, reducing the per-iteration complexity from cubic to linear in the number of antennas and the initialization from a matrix inversion to a matched filter, while retaining competitive \ac{BER} performance.}
\item We introduce a principled method for selecting the regularization weight that trades data fidelity against the discreteness penalty, based on Morozov's discrepancy principle, which eliminates the need for hand-tuning or cross-validation.
\item {Through extensive simulations across a range of system dimensions, \ac{DAC} resolutions, and channel impairments---including imperfect \ac{CSI} and spatial correlation---we demonstrate that both proposed algorithms achieve competitive or superior \ac{BER} performance compared to state-of-the-art quantized precoders, while the \ac{GaBP} variant attains this at linear complexity.}
\end{enumerate}

\begin{figure*}[tp]
\centering
\includegraphics[width=\textwidth]{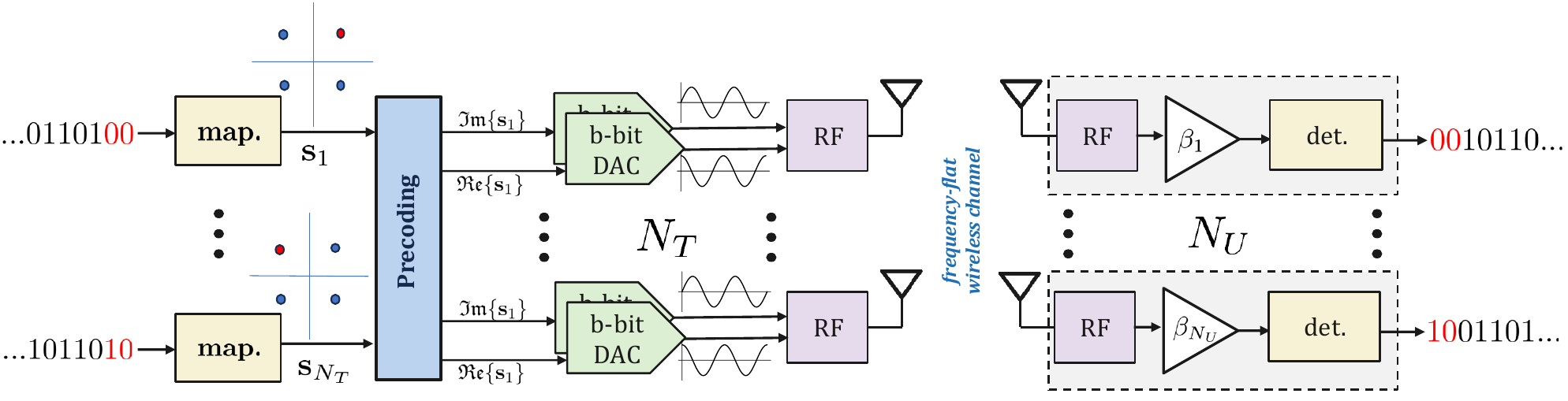}
\caption{System model of the downlink of a \ac{MU-mMIMO} system with low-resolution \acp{DAC}, where the \ac{BS} is equipped with $N_T$ antennas and serves $N_U$ single-antenna users.
}
\label{fig:systemmodel}
\end{figure*}

\vspace{-3ex}
\subsection{Notation}

Boldface lowercase letters (e.g., $\mathbf{x}$) denote column vectors and boldface uppercase letters (e.g., $\mathbf{H}$) denote matrices; non-bold letters denote scalars.
The $i$th entry of $\mathbf{x}$ is written $x_i$, the $(i,j)$th entry of $\mathbf{H}$ is $[\mathbf{H}]_{i,j}$, and $\mathbf{h}_u$ denotes the $u$th row of $\mathbf{H}$.
$\|\mathbf{x}\|_2$ is the Euclidean ($\ell_2$) norm of $\mathbf{x}$, $\mathbf{I}_N$ is the $N\times N$ identity matrix, and $\mathrm{diag}(\mathbf{x})$ is the diagonal matrix formed from $\mathbf{x}$.
The superscripts $(\cdot)^{\mathrm{T}}$, $(\cdot)^{*}$, and $(\cdot)^{\mathrm{H}}$ denote transpose, complex conjugate, and Hermitian (conjugate) transpose, respectively.
$\mathbb{R}$ and $\mathbb{C}$ denote the sets of real and complex numbers; $\mathrm{j}\triangleq\sqrt{-1}$ is the imaginary unit.
$\Re\{\cdot\}$ and $\Im\{\cdot\}$ extract the real and imaginary parts; $|\cdot|$ denotes the absolute value (or cardinality of a set).
$\mathbb{E}[\cdot]$ denotes expectation, and $\mathbf{z}\sim\mathcal{CN}(\mathbf{0},\sigma^2\mathbf{I}_N)$ denotes a circularly-symmetric complex Gaussian vector with zero mean and covariance $\sigma^2\mathbf{I}_N$.
Finally, $\mathcal{C}_b$ denotes the per-dimension quantization alphabet of a $b$-bit \ac{DAC}, so that the admissible transmit set is $\mathcal{C}_b^{N_T}$, and $\mathcal{Q}_b(\cdot)$ denotes the element-wise quantization (mapping) onto this alphabet.

\vspace{-2ex}
\subsection{Paper Outline}
%
% --- Outline ---
The rest of this paper is organized as follows.
Section~\ref{sec:system} introduces the quantized multi-user \ac{MIMO} downlink system model and formalizes the receive-side \ac{MSE} design criterion under a finite-resolution \ac{DAC} constraint.
Section~\ref{sec:Formulation} reviews the conventional surrogate formulation commonly adopted for one-bit quantized precoding and discusses its limitations at higher resolutions; {it then introduces the proposed discrete-aware \ac{MSE} formulation, along with the formal derivations and definitions.}
Section~\ref{sec:message_passing} derives the second proposed algorithm, based on \ac{GaBP}, and details its message-passing schedule and complexity.
Section~\ref{sec:Simulations} reports numerical results comparing the proposed methods against infinite-resolution and state-of-the-art quantized baselines, across different system configurations.
Finally, Section~\ref{sec:conclusion} concludes the paper.

\section{System Model and Nonlinear Precoding}
\label{sec:system}
\subsection{System Model}

{We consider the downlink of a single-cell \ac{MU-mMIMO} system with a \ac{BS} equipped with $N_T$ antennas serving $N_U$ single-antenna users over the same time-frequency resources}, as illustrated in Fig. \ref{fig:systemmodel}.
{Assuming narrowband signaling and perfect synchronization, we model the received signal at the~$N_U$ users  as follows:}
\begin{equation}
\mathbf{y} = \mathbf{H} \mathbf{x} + \mathbf{n} \in \mathbb{C}^{N_U \times 1}.
\label{eq:rx}
\end{equation}

{Here, $\mathbf{H} \in \mathbb{C}^{N_U \times N_T}$ models the downlink channel between the \ac{BS} and the $N_U$ users; the entry $[\mathbf{H}]_{u,t}$ represents the complex channel gain between the $t$th \ac{BS} antenna and the $u$th user, for $u = 1, \dots, N_U$ and $t = 1, \dots, N_T$, and is assumed to be perfectly known at the \ac{BS}}\footnote{The impact of imperfect \ac{CSI} is investigated in Sec.~\ref{sec:Simulations}.}.
{Finally, the entries of the noise vector $\mathbf{n}$ are drawn from a circularly-symmetric complex Gaussian process with per-entry variance $\sigma_n^2$, which is also assumed to be known perfectly at the \ac{BS}.}
{We denote the transmitted, precoded, and quantized signal by $\mathbf{x} \in \mathcal{C}_b^{N_T}$, where $\mathcal{C}_b^{N_T}$ is the alphabet of a $b$-bit \ac{DAC} across the $N_T$ \ac{BS} antennas.}

As illustrated in Fig.~\ref{fig:systemmodel}, each of the $N_T$ transmit antennas is equipped with a \emph{pair} of \acp{DAC}, one driving the in-phase (real) and one the quadrature (imaginary) branch, and each \ac{DAC} is characterized by its own set of (real-valued) quantization levels.
These quantization levels are collected in the set
$\mathcal{L} = \{\ell_0, \dots, \ell_{L-1}\}$, with
$L = |\mathcal{L}|$ levels per real dimension.
{For convenience, the number of quantization bits per real dimension is defined as $b = \log_2(|\mathcal{L}|) = \log_2 L$.}
Because each of the $N_T$ antennas at
the transmitter uses the same $b$-bit
quantization alphabet, the per-antenna \emph{complex} alphabet is built from its two real branches as $\mathcal{C}_b = \mathcal{L} \times \mathcal{L}$, and the overall transmit alphabet is the $N_T$-fold product $\mathcal{C}_b^{N_T} = (\mathcal{L}\times\mathcal{L})^{N_T}$.
For any finite resolution $b$, the alphabet
$\mathcal{C}_b^{N_T}$ is a finite set and can therefore
never equal the uncountable field $\mathbb{C}^{N_T}$. Nevertheless, as the
resolution grows ($b \to \infty$), the levels in $\mathcal{L}$ become
arbitrarily dense and $\mathcal{C}_b$ becomes dense in $\mathbb{C}$.
Accordingly, the infinite-resolution (unquantized) case is {modeled} by
relaxing the transmit alphabet to $\mathbf{x} \in \mathbb{C}^{N_T}$, which
serves as a limiting reference rather than as an attainable alphabet.
{As shown later, this relaxation serves as a
lower bound on error-rate performance.}

\subsection{Uniform Quantization Model}
\label{sec:quant}

{The quantization performed by the \acp{DAC} is modeled as a mid-rise uniform quantizer, which maps the input signal to the nearest level in the set $\mathcal{L}$, with each quantization level given by}
\begin{equation}
\ell_l = \alpha \left(l - \frac{|\mathcal{L}|-1}{2}\right) \Delta, \quad l = 0, 1, \dots, |\mathcal{L}|-1 .
\end{equation}

{For a mid-rise $b$-bit quantizer, the quantization levels are uniformly spaced with step size $\Delta$, which is designed to minimize the quantization error.}
{In addition, the levels are symmetric around zero, as is common in the literature, and a scaling factor $\alpha$ is introduced to satisfy the power constraint.}

{Similarly, we define the quantization thresholds as}
\begin{equation}
\tau_l = \alpha \left(l - \frac{|\mathcal{L}|}{2}\right) \Delta, \quad l = 1, \dots, |\mathcal{L}|-1,
\end{equation}
with $\tau_0 = -\infty$ and $\tau_{|\mathcal{L}|} = \infty$, which can be grouped into the set $\mathcal{T} = \{\tau_0, \tau_1, \dots, \tau_{|\mathcal{L}|}\}$.

Together, the quantization levels and thresholds define the $b$-bit quantization function $\mathcal{Q}_b(\cdot)$, which maps a real-valued input to the nearest quantization level in $\mathcal{L}$,
\begin{equation}
\mathcal{Q}_b(x) = \ell_l \quad \text{if} \quad x \in [\tau_l, \tau_{l+1}),
\end{equation}
which can also be applied element-wise to a complex-valued input by quantizing separately to the real and imaginary parts, so that the resulting per-antenna symbol lies in $\mathcal{C}_b=\mathcal{L}\times\mathcal{L}$.

\subsection{Nonlinear Precoding}
\label{sec:precoding}

{We assume that the symbols to be transmitted are collected in the vector $\mathbf{s}\in\mathcal{S}^{N_U}$, drawn from a discrete constellation $\mathcal{S}$ (e.g., QAM or PSK). The \ac{BS} uses the \ac{CSI} $\mathbf{H}$ to compute the precoded and quantized signal $\mathbf{x}$ transmitted over the wireless channel.}
As noted in \cite{Jacobsson2017quantized}, the function $\mathcal{P}$ that maps the input symbol vector $\mathbf{s} \in \mathcal{S}^{N_U}$ to the precoded and quantized signal $\mathbf{x} \in \mathcal{C}_b^{N_T}$ is, in general, nonlinear, i.e., $\mathbf{x} = \mathcal{P}(\mathbf{s}, \mathbf{H})$, which is commonly referred to as nonlinear precoding.

{The transmit signal $\mathbf{x}$ must further satisfy the power constraint}\footnote{This deterministic power constraint on $\mathbf{x}$ could instead be imposed in average terms, but is, as in \cite{Jacobsson2017quantized}, adopted here in order to allow for a convenient problem formulation.}
\begin{equation}
    \|\mathbf{x}\|_2^2 \leq P.
    \label{eq:power_constraint}
\end{equation}

For a given symbol vector $\mathbf{s}$ and power constraint,
our goal is to design a precoding vector $\mathbf{x}$
and a constant $\beta\in\mathbb{R}$ that minimize the receive-side \ac{MSE} given by \cite{Jacobsson2017quantized}
\begin{align}
\mathbb{E}_\mathbf{n}\!\left[\,\|\mathbf{s} - \beta\mathbf{y}\|_2^2\,\right] = \|\mathbf{s} - \beta\mathbf{H}\mathbf{x}\|_2^2 + \beta^2 {N_U \sigma_n^2} .
\end{align}

Here, $\beta$ is a common scaling factor applied at the receivers, so that each $u$th
\ac{UE} forms its estimate as $\hat{s}_u = \beta y_u$.
{We therefore design the transmit
vector $\mathbf{x}$ so that each $u$th \ac{UE} recovers its intended symbol $s_u$, up to the common factor $\beta$, from
$\hat{s}_u \triangleq \beta y_u = \beta\!\left(\mathbf{h}_u\mathbf{x} + n_u\right)$, where~$\mathbf{h}_u$ denotes the $u$th row of~$\mathbf{H}$. This design attains the smallest possible \ac{MSE}
with respect to $\mathbf{s}$ while respecting the power budget
in \eqref{eq:power_constraint}.}
With infinite-resolution \acp{DAC}, $\mathbf{x}$ may take any value in
$\mathbb{C}^{N_T}$, which is by now well investigated \cite{Christensen2008WMMSE}.
{Finite-resolution \acp{DAC}, in contrast, confine $\mathbf{x}$ to the discrete set
$\mathcal{C}_b^{N_T}$. Only finitely many transmit vectors are then admissible, and the quantization introduces unavoidable distortion at the receivers.}
Minimizing this \ac{MSE} under the discrete constraint is therefore a combinatorial problem
for which an exactly \ac{MMSE}-optimal solution is generally out of reach.
We therefore focus on low-complexity algorithms that approximate this solution.

{The next sections review the commonly used surrogate formulation for one-bit quantized
precoding and then present the proposed discrete-aware formulation, which applies to
arbitrary quantization alphabets, along with efficient algorithms.}
\section{Problem Formulation}
\label{sec:Formulation}

{Next, we present the problem formulation for the design of the quantized and precoded signal $\mathbf{x}$.}
{As explained in the previous section, in the \ac{MMSE} sense, we design the optimal signal $\mathbf{x}$ by solving the following optimization problem \cite{Jacobsson2017quantized}} 
\begin{align}
\operatorname*{minimize}_{\mathbf{x}\in\mathbb{C}^{N_T \times 1}, \beta \in \mathbb{R}^+} \quad & \|\mathbf{s} - \beta\mathbf{H} \mathbf{x}\|_2^2 + \beta^2 {N_U \sigma_n^2}\label{eq:mmse} \\
\text{subject to} \quad & \|\mathbf{x}\|_2^2 \leq P.\nonumber
\end{align}

After solving the optimization problem, the \ac{BS} transmits $\mathbf{x}$ over the channel, and the users scale their received signals in $\mathbf{y}=\mathbf{H}\mathbf{x}+\mathbf{n}$ to obtain estimates of the transmitted signals as
\begin{equation}
\hat{\mathbf{s}} = \beta\mathbf{y} = \beta (\mathbf{H} \mathbf{x} + \mathbf{n}).
\end{equation}
Here, $\hat{\mathbf{s}}$ collects the estimates $\hat{s}$ defined above.

{The optimization problem in \eqref{eq:mmse} is non-convex and combinatorial, thus \ac{NP}-hard in general.}
{Although the optimal $\beta$ can be computed in closed form, for a fixed $\mathbf{x}$, via \cite{Jacobsson2017quantized}}
%\cs{same P normalization issue}
%
\begin{equation}
\label{eq:beta_opt}
\beta^\star = \frac{\Re\{\mathbf{s}^H \mathbf{H} \mathbf{x}\}}{\|\mathbf{H} \mathbf{x}\|_2^2 + {N_U \sigma_n^2}},
\end{equation}
{it is the joint optimization of both variables, together with the discrete constraint on $\mathbf{x}$, that makes the problem challenging.}

{We first review the \ac{SotA} formulation for one-bit quantized precoding, then present the proposed discrete-aware formulation.}

\subsection{{SotA} Formulations}
\label{sec:sota}

{Noting that all one-bit \ac{DAC} outputs share the same magnitude, and setting $\alpha \Delta = \sqrt{\frac{2P}{N_T}}$, the authors of \cite{Jacobsson2017quantized} proposed to solve the following optimization problem} 
\begin{align}
\operatorname*{minimize}_{\mathbf{b} \in \mathcal{B}^{N_T}} \quad & \|\mathbf{s} - \mathbf{H} \mathbf{b}\|_2^2 + \frac{N_U \sigma_n^2}{P}\|\mathbf{b}\|_2^2 \label{eq:jacobsson}
\end{align}
where $\mathcal{B} = \left\{\sqrt{\frac{P}{2N_T}}(\pm \chi \pm j\chi), \forall \chi > 0\right\}$ is the set of possible outputs of the one-bit \ac{DAC}.
The problem in \eqref{eq:jacobsson} is a surrogate formulation for one-bit quantized precoding, which
transforms the discrete constraint on $\mathbf{x}$ by allowing it to take values in the set $\mathcal{B}$, which is the union of the four half-lines emanating from the origin through the one-bit corner points.
{On this basis, the authors of  \cite{Jacobsson2017quantized}  derived several algorithms to solve the problem, such as  semidefinite relaxation (SDR) and its efficient version, the squared-infinity norm relaxation~(SQUID).}

{In contrast to the proposed formulation, the linear precoders explained below are subject to an average transmit-power constraint and do not account for the quantization distortion, therefore serving as baselines for comparison.}
\subsubsection{WF precoding} For {infinite-resolution} \acp{DAC}, the optimal {linear} precoder is the 
\ac{MMSE} (or \ac{WF}), which minimizes the \ac{MSE} between the transmitted and received signals, given by
\begin{equation}
\label{eq:MMSE}
\mathbf{x}^{\text{MMSE}} = \frac{1}{\rho^{\text{MMSE}} }\mathbf{H}^H (\mathbf{H} \mathbf{H}^H + \frac{N_U \sigma_n^2}{P} \mathbf{I})^{-1} \mathbf{s},
\end{equation}
with the optimum power scaling factor $\rho$ given by 
\begin{equation}
\rho^{\text{MMSE}} =\! \frac{1}{\sqrt{P}}\text{tr}\bigg(\!\Big(\mathbf{H} \mathbf{H}^{\herm}\!\! + \tfrac{N_U \sigma_n^2}{P} \mathbf{I}\Big)^{\!\!-1}
\!\! \mathbf{H} \mathbf{H}^{\herm} \Big(\mathbf{H}\mathbf{H}^{\herm} \!\! + \tfrac{N_U \sigma_n^2}{P} \mathbf{I}\Big)^{\!-1}\bigg)^{\!\!\frac{1}{2}}\!\!.
\end{equation}

\subsubsection{MRT precoding} The \ac{MRT} precoder is a simple linear precoder that maximizes the signal power at the receiver, and is given by
\begin{equation}
\label{eq:MRT}
\mathbf{x}^{\text{MRT}} = \frac{1}{\rho^{\text{MRT}}} \mathbf{H}^H \mathbf{s},
\end{equation}
with the optimum scaling factor $\rho$ given by
\begin{equation}
\rho^{\text{MRT}} = \frac{1}{N_T\sqrt{P}}\sqrt{\text{tr}\Bigl(\mathbf{H} \mathbf{H}^{\herm}\Bigr)}.
\end{equation}

\subsubsection{ZF precoding} Similarly, the \ac{ZF} precoder is a simple linear precoder that cancels the interference at the receiver, and is described by
\begin{equation}
\label{eq:ZF}
\mathbf{x}^{\text{ZF}} = \frac{1}{\rho^{\text{ZF}}} \mathbf{H}^{\herm}(\mathbf{H} \mathbf{H}^{\herm})^{-1} \mathbf{s},
\end{equation}
with the optimum scaling factor $\rho$ given by
\begin{equation}
\rho^{\text{ZF}} = \frac{1}{\sqrt{P}}\sqrt{\text{tr}\Bigl((\mathbf{H} \mathbf{H}^{\herm})^{-1}\Bigr)}.
\end{equation}

{For a $b$-bit quantizer, we follow the linear precoders by an element-wise quantization to the nearest constellation point in~$\mathcal{C}_b^{N_T}$, giving the quantized linear precoders}
\begin{align}
\mathbf{x}^{\text{Precod-Quant}} &= \mathcal{Q}_b(\mathbf{x}^{\text{Precoded}}).
\end{align}

\begin{algorithm}[H]
\caption{\ac{CDM} Quantized Precoding}
\label{alg:cdm}
\begin{algorithmic}[1]
\State \textbf{Input:} symbols $\mathbf{s}$, channel $\mathbf{H}$, alphabet $\mathcal{C}_b$, iterations $T_{\max}$
\State \textbf{Init:} $\mathbf{x} \gets \mathcal{Q}_b(\mathbf{W}_{\mathrm{MMSE}}\,\mathbf{s})$,\;\; $\mathbf{z} \gets \mathbf{H}\mathbf{x}$,\;\; $k \gets 0$
\Repeat
\For{$n = 1$ \textbf{to} $N_T$}
    \State Compute $c^{\star} \gets \displaystyle\arg\min_{c \in \mathcal{C}_b} \; J\!\big(\mathbf{z} + (c - x_n)\,\mathbf{h}_n\big)$
    \State Update $\mathbf{z} \gets \mathbf{z} + (c^{\star} - x_n)\,\mathbf{h}_n$,
    \State Assign $x_n \gets c^{\star}$
\EndFor
\State $k \gets k + 1$
\Until{$k > T_{\max}$ \textbf{or} converged}
\State \Return $\mathbf{x}$
\end{algorithmic}
\end{algorithm}
\vspace{-2ex}

\subsubsection{Coordinate descent precoding}
{Beyond the linear baselines, we also consider a precoder based on the coordinate-descent method (\ac{CDM}), originally developed for the equalization (data-detection) problem~\cite{Chen2022OneBitMMSE} and adapted here to the precoding setting.}
Initialized from a quantized \ac{MMSE} solution, it greedily revisits one antenna at a time and assigns to it the alphabet symbol that minimizes the precoding objective $J(\cdot)$, which coincides with \eqref{eq:mmse} for fixed $\beta$.

In CDM, each candidate is scored at a low cost through a rank-one update of the noiseless received vector $\mathbf{z}=\mathbf{H}\mathbf{x}$, and the sweep is repeated until convergence, as summarized above in Algorithm~\ref{alg:cdm}.

\vspace{-1ex}
\subsection{Proposed Discrete-Aware Formulation}
\label{sec:proposed}

{Although the reformulation in \eqref{eq:jacobsson} is a good surrogate for one-bit quantized precoding, it does not explicitly enforce the discrete constraint on $\mathbf{x}$, which renders it inapplicable to higher-resolution quantization.}
{In addition, methods that attempt to find solutions require matrix inversions, which can  computationally expensive for large systems.}
{In contrast, the proposed formulation applies to arbitrary quantization resolutions and can be solved efficiently.}
{However, since this problem must be solved for each symbol vector $\mathbf{s}$, the computational complexity of the algorithm is a critical consideration.}
{To this end, we also present a message-passing algorithm that solves the proposed formulation with linear complexity in the number of antennas, a significant improvement over all previous formulations.}

{In order to explicitly enforce the discrete constraint on $\mathbf{x}$, we utilize an
$\ell_0$-norm penalty, which counts the number of non-zero entries in a vector, within the
optimization problem in~\eqref{eq:mmse}. The use of such a penalty to impose discrete,
finite-alphabet constraints was originally proposed in the context of \ac{MIMO}
data detection~\cite{Iimori2021Robust}; here, we borrow the principle for quantized precoding.}
{Since the transmit-power constraint is active at
the optimum---the full power budget is always used to minimize the \ac{MSE}---we have
$\|\mathbf{x}\|_2^2 = P$. Substituting this identity into the noise term, i.e.,
$\beta^2 N_U\sigma_n^2 = \beta^2\tfrac{N_U\sigma_n^2}{P}\|\mathbf{x}\|_2^2$, renders the
objective scale-invariant in $\mathbf{x}$; the explicit power constraint thus becomes can be relaxed, leaving only the discrete-alphabet constraint}
\begin{align}
\operatorname*{minimize}_{\mathbf{x}\in \mathbb{C}^{N_T}, \beta \in \mathbb{R}^+} \quad & \|\mathbf{s} - \beta\mathbf{H} \mathbf{x}\|_2^2 + \beta^2\frac{N_U \sigma_n^2}{P}\|\mathbf{x}\|_2^2 \label{eq:l0} \\[-1ex]
\text{subject to} \quad & \sum_{i=1}^{2^{2b}} \|\mathbf{x} - c_i \mathbf{1}\|_0 = N_T (2^{2b} - 1). \nonumber
\end{align}

{Here, $c_i$ denotes the $i$th constellation point in the quantization alphabet, onto which every entry of $\mathbf{x}$ is forced via the constraint.}
{The intuition is as follows: the term $\|\mathbf{x} - c_i \mathbf{1}\|_0$ counts how many entries of $\mathbf{x}$ differ from $c_i$. Since each of the $N_T$ entries can match only one of the $2^{2b}$ levels, it differs from the remaining $2^{2b}-1$. The total count therefore equals $N_T (2^{2b} - 1)$ exactly when every entry lies in the alphabet.}
{Owing to this scale invariance, fixing $\beta$ to its optimal value incurs no loss.
From now on we therefore focus on optimizing $\mathbf{x}$, with the optimal scaling given in
closed form by~\eqref{eq:beta_opt} and denoted $\beta^\star(\mathbf{x})$.}

{Hereafter, we focus on optimizing $\mathbf{x}$ and assume the optimal $\beta$ to be given by \eqref{eq:beta_opt}, denoted $\beta^\star(\mathbf{x})$.}
{We then reformulate \eqref{eq:l0} into a convex problem that can be solved efficiently, and derive a closed-form solution that we update iteratively until convergence.}
{To that end, start by introducing a regularization parameter $\lm$ into the optimization problem in \eqref{eq:l0}, which, for a fixed $\beta^\star(\mathbf{x})$, gives}\footnote{For the sake of notational simplicity, we omit hereafter $\beta^\star$'s explicit dependence on $\mathbf{x}$.}
\vspace{-2ex}
\begin{equation}
\operatorname*{minimize}_{\mathbf{x} \in \mathbb{C}^{N_T}} \,\, 
\|\mathbf{s} - \beta^\star \mathbf{H} \mathbf{x}\|_2^2
+ {\beta^\star}^2 \frac{N_U\sigma_n^2}{P}\|\mathbf{x}\|_2^2 + \lm \sum_{i=1}^{2^{2b}} \|\mathbf{x} - c_i \mathbf{1}\|_0.
\label{eq:l0_reg}
\end{equation}

{The regularization parameter $\lm\geq0$ controls the trade-off between the data-fidelity term and the discrete constraint. We exploit this trade-off in the subsequent steps to derive an efficient algorithm.}
{And since the last term in the objective is the non-convex, non-smooth $\ell_0$-norm, we first apply a smooth convex approximation to it, which lets us reformulate the problem into one that can be solved efficiently.}

\subsubsection{Convex Smooth Approximation of the \texorpdfstring{$\ell_0$}{l0} Norm}
{We approximate the \texorpdfstring{$\ell_0$}{l0}-norm by a smooth function, such as the log-sum-exp or quadratic function.}
{In this work, we use the following quadratic approximation of the $\ell_0$-norm:}
\begin{equation}
\|\mathbf{z}\|_0 \approx \sum_{m=1}^{M} \frac{|z_m|^2}{|z_m|^2 + \zeta} = M - \sum_{m=1}^{M} \frac{\zeta}{|z_m|^2 + \zeta},
\label{eq:l0_approx}
\end{equation}
which {becomes exact as $\zeta \to 0$.}

{Substituting \eqref{eq:l0_approx} into \eqref{eq:l0_reg} and dropping constant terms that do not affect the minimizer yields}
\begin{align}
\label{eq:smooth}
\operatorname*{minimize}_{\mathbf{x} \in \mathbb{C}^{N_T}} \,\, &   \|\mathbf{s}\! -\! \beta^\star \mathbf{H} \mathbf{x}\|_2^2 + \!{\beta^\star}^2 \frac{N_U\sigma_n^2}{P}\|\mathbf{x}\|_2^2 \notag \\[-1ex]
& -\lm \sum_{i=1}^{2^{2b}} \sum_{j=1}^{N_T} \frac{\zeta}{|x_j\! -\! c_i|^2\! +\! \zeta}.
\end{align}

{The problem in \eqref{eq:smooth} is still non-convex, but now smooth. To solve it efficiently, we only need to convexify the last term of the objective, which we achieve with the \ac{FP} approach in the next section.}

\subsubsection{Fractional Programming convexification}
{Using the \ac{FP} approach \cite{Shen2018}, we convexify the last term in \eqref{eq:smooth} by introducing the auxiliary variables $\gamma_{i,j}$, given by}
\begin{equation}
\gamma_{i,j} = \frac{\sqrt{\zeta}}{|x_j - c_i|^2 + \zeta},
\label{eq:gamma}
\end{equation}
{which we update in each iteration of the algorithm.}

{In light of the above, the $\ell_0$-norm approximation  becomes}
\begin{align}
\|\mathbf{z}\|_0
  &\approx M - \Bigl(\sum_{m=1}^{M} 2\gamma_m \sqrt{\zeta}
        - \gamma_m^2 \bigl(|z_m|^2 + \zeta\bigr)\Bigr) \nonumber\\[-0.5ex]
  &= \underbrace{M - \sum_{m=1}^{M}\bigl(2\gamma_m \sqrt{\zeta}
        - \gamma_m^2 \zeta\bigr)}_{\text{const.\ w.r.t.\ }\mathbf{z}}
        + \sum_{m=1}^{M} \gamma_m^2\,|z_m|^2 .
\end{align}

{With this approximation, and holding the weights $\gamma_{i,j}$ fixed at their current values, we express the problem in \eqref{eq:smooth} as a smooth, convex quadratic problem}
\begin{align}
\operatorname*{minimize}_{\mathbf{x} \in \mathbb{C}^{N_T}} \,\, & \|\mathbf{s} - \beta^\star \mathbf{H} \mathbf{x}\|_2^2 + {\beta^\star}^2 \frac{N_U\sigma_n^2}{P} \|\mathbf{x}\|_2^2 \nonumber \\
& \quad + \lm \sum_{i=1}^{2^{2b}} \sum_{j=1}^{N_T} \gamma_{i,j}^2 |x_j - c_i|^2. \label{eq:quadratic}
\end{align}

Since $\gamma_{i,j}$ in fact depend on $\mathbf{x}$, \eqref{eq:quadratic} is a surrogate that coincides with \eqref{eq:smooth} only at a fixed point; we therefore hold the weights fixed within each iteration and update them in between.
{Following the same steps as in \cite{Iimori2021Robust}, we collect the terms in the last summation into a more compact form, from which we derive a closed-form solution for $\mathbf{x}$.}

{First, we define the following quantities:}
\begin{align}
\mathbf{g} & = \sum_{i=1}^{2^{2b}} c_i [\gamma_{i,1}^2, \dots, \gamma_{i,N_T}^2]\trans \in \mathbb{C}^{N_T\times 1} \label{eq:g} \\
\mathbf{G} & = \sum_{i=1}^{2^{2b}} \operatorname{diag}(\gamma_{i,1}^2, \dots, \gamma_{i,N_T}^2) \succ 0. \label{eq:G}
\end{align}

The optimization problem is then equivalent to
\begin{align}
\operatorname*{minimize}_{\mathbf{x} \in \mathbb{C}^{N_T}} \; &\|\mathbf{s} - \beta^\star \mathbf{H} \mathbf{x}\|_2^2 + {\beta^\star}^2\frac{N_U\sigma_n^2}{P}\|\mathbf{x}\|_2^2  \notag \\
&  +\lm (\mathbf{x}^H \mathbf{G} \mathbf{x} - 2 \Re\{\mathbf{g}^H \mathbf{x}\}),
\label{eq:elegant}
\end{align}
{which is the main focus of the analysis from now on: a quadratic problem with a closed-form solution that we update iteratively until convergence.}

\subsection{Closed-Form Solution}

{Before deriving the closed-form solution for $\mathbf{x}$, we first renormalize the problem by absorbing the scaling factor $\beta^\star$ into the symbol vector $\mathbf{s}$ and the noise term, which gives}
\begin{equation}
\operatorname*{minimize}_{\mathbf{x} \in \mathbb{C}^{N_T}} \,\,   \|\check{\mathbf{s}} - \mathbf{H} \mathbf{x}\|_2^2 +\frac{N_U\sigma_n^2}{P}\|\mathbf{x}\|_2^2  + \check{\lm} (\mathbf{x}^H \mathbf{G} \mathbf{x} - 2 \Re\{\mathbf{g}^H \mathbf{x}\}),
\label{eq:renormalized}
\end{equation}
in which $\check{\mathbf{s}} = \mathbf{s}/\beta^\star$ and $\check{\lm} = \lm/(\beta^\star)^2$.

{Since the above problem is quadratic, we derive a closed-form solution for $\mathbf{x}$ by taking the Wirtinger derivative of the objective with respect to $\mathbf{x}$ and setting it to zero}
\begin{equation}
\frac{\partial}{\partial \mathbf{x}^*} \Bigl( \|\check{\mathbf{s}} - \textstyle \mathbf{H} \mathbf{x}\|_2^2 +\frac{N_U\sigma_n^2}{P}\|\mathbf{x}\|_2^2  + \check{\lm} (\mathbf{x}^H \mathbf{G} \mathbf{x} - 2 \Re\{\mathbf{g}^H \mathbf{x}\}) \! \Bigr) \! = \! 0,
\end{equation}
which after some algebraic manipulations yields the following closed-form solution for $\mathbf{x}$
\begin{equation}
\mathbf{x} = \left( \mathbf{H}^H \mathbf{H} + \check{\lm} \mathbf{G} + \frac{N_U \sigma_n^2}{P} \mathbf{I} \right)^{\!-1} (\mathbf{H}^H \check{\mathbf{s}} + \check{\lm} \mathbf{g}).
\label{eq:closed_mmse}
\end{equation}

{Notice that the above expression for $\mathbf{x}$ is a least-squares-type solution, similar to the 
%\cs{linear? }
{linear} \ac{MMSE} precoder but with an additional regularization term $\check{\lm} \mathbf{G}$ that accounts for the discrete constraint on $\mathbf{x}$.}

\subsection{Principled Selection of the Regularization Weight}
\label{sec:lambda}

The weight $\lambda$ in \eqref{eq:l0_reg} balances data fidelity against the
discreteness penalty, and the recovered precoder is sensitive to its value.
{Instead of tuning it by grid search or cross-validation, which is costly and difficult to adapt to different propagation conditions and operating \ac{SNR}, we fix it from a classical principle: the data residual should be driven down to, but not below, the floor set by the thermal noise.}
This is \emph{Morozov's discrepancy principle}, a standard {\emph{a~posteriori}} rule for {selecting the} regularization weight in inverse problems \cite{Morozov1984, EnglHankeNeubauer1996}.

Let
$\mathbf{x}(\check\lambda)$ denote the closed form \eqref{eq:closed_mmse} for the
current \ac{FP} weights $(\mathbf{G},\mathbf{g})$, with renormalized data misfit
$F(\check\lambda)=\|\check{\mathbf{s}}-\mathbf{H}\mathbf{x}(\check\lambda)\|_2^2$.
No precoder can drive this below the distortion injected by the noise, whose level is on average
$\delta^\star = N_U\sigma_n^2/P$, {so we select $\check\lambda$ such that the misfit matches this floor}, $F(\check\lambda)=\delta^\star$.
Since $F$ increases monotonically with $\check\lambda$, this root is unique.

{A high-accuracy root search is unnecessary inside an already iterative scheme, so we take a single damped, log-domain step per iteration, letting $\check\lambda$ converge
alongside the \ac{FP} iterates,}
\begin{equation}
\check\lambda \;\leftarrow\;
    \Pi_{[\lambda_{\min},\,\lambda_{\max}]}\!
    \left(\check\lambda\cdot\Big(\frac{\delta^\star}{F(\check\lambda)}\Big)^{\!\kappa}\right),
\label{eq:lambda_update}
\end{equation}
where $\Pi_{[\cdot]}$ projects onto
$[\lambda_{\min},\lambda_{\max}]$ for numerical safety and $\kappa\in(0,1]$ is a step gain\footnote{Notice that $\lambda_{\min}$, $\lambda_{\max}$, and $\kappa$ are not hyperparameters that must be tuned, but merely fixed constants typically fixed heuristically for numerical efficiency/stability purposes. The interval $[\lambda_{\min},\lambda_{\max}]$, in particular, is only a loose numerical clamp that keeps $\check\lambda$ from underflowing to $0$ or diverging during the iteration;}.

The step raises $\check\lambda$ when $F<\delta^\star$ and lowers it when $F>\delta^\star$. As $\mathbf{x}(\check\lambda)$ is already formed by \eqref{eq:closed_mmse}, the update reuses the residual it produces and adds only a few scalar operations, leaving no weight to hand-tune.
{We then enforce discreteness through the \emph{shape} of the penalty rather than by inflating the fidelity target: we anneal the smoothing parameter $\zeta$ in \eqref{eq:gamma} from a broad $\zeta_0$ down to a small $\zeta_{\min}$ over the iterations via graduated non-convexity}
\begin{equation}
\zeta^{(t)} = \zeta_0\,\big(\zeta_{\min}/\zeta_0\big)^{\min\{1,\,t/T_{\max}\}},
\label{eq:anneal}
\end{equation}
so that early iterations optimize a nearly convex surrogate and later ones sharpen the penalty wells onto the alphabet $\mathcal{C}_b$.

{We do not claim that this is the only or best way to select $\check\lambda$; rather, it is a principled, low-complexity approach that performs well in practice.}
For more advanced selection rules, one can refer to \cite{Boukari1995, Pedregosa2016, Bischl2023, Ehrhardt2024} and references therein.

\subsection{Proposed Algorithm}

{We now summarize the proposed algorithm for solving the optimization problem in \eqref{eq:l0}, referred to as the \ac{MQP} algorithm.}
{The proposed algorithm iteratively updates the auxiliary \ac{FP} variables $\gamma_{i,j}$, the regularization parameter $\lambda$, the precoded signal $\mathbf{x}$, and $\beta^\star$ until convergence.}

\begin{algorithm}[H]
\caption{\ac{MQP}}
\label{alg:MQP}
\begin{algorithmic}[1]
\State \textbf{Input} $\bm{s}$, $\mathbf{H}$, $\sigma_{\bm{n}}^2$, $\zeta$, $\mathcal{C}_b^{N_T}$
\Repeat
\State Compute $\gamma_{i,j}$ as in \eqref{eq:gamma}
\State Compute $\mathbf{g}$ and $\mathbf{G}$ as in \eqref{eq:g} and \eqref{eq:G}
\State Update $\mathbf{x}$ using \eqref{eq:closed_mmse}
\State Update $\lambda$ from the residual using \eqref{eq:lambda_update}
\State Update $\beta^\star$ using \eqref{eq:beta_opt}
\Until{convergence}
\State \textbf{Output} $\mathcal{Q}_b(\mathbf{x}^\star)$
\end{algorithmic}
\end{algorithm}

Most of the complexity of the algorithm stems from the matrix inversion in the closed-form solution for $\mathbf{x}$, which
must be recomputed at every outer iteration, since $\mathbf{G}$ and $\check{\lambda}$ change between iterations.
Its cost can be \emph{reduced}---though not eliminated---by means of the Woodbury matrix identity: exploiting that $\mathbf{G}$ is diagonal and that the channel Gram $\mathbf{H}^H\mathbf{H}$ has rank at most $N_U$, the inversion can be recast so that the only non-trivial inverse is of size $N_U\times N_U$.
Without stating the explicit expression, this lowers the per-iteration cost to at most $\mathcal{O}(N_U^2 N_T)$, i.e., linear in the number of antennas $N_T$ and quadratic only in the much smaller number of users $N_U$.
{Even so, this cost remains non-negligible; the next section shows how a message-passing formulation reduces it further.}

Unlike eigenvalue- or search-based selection rules, the discrepancy-principle update in \eqref{eq:lambda_update} adds only a few scalar operations per iteration, since it reuses the residual already formed for the closed-form update in \eqref{eq:closed_mmse}.
The per-iteration complexity is therefore dominated by the single linear solve, with no auxiliary optimization problem and no offline look-up table required.
{Algorithm~\ref{alg:MQP} summarizes the proposed method.}

\section{Message Passing Formulation}
\label{sec:message_passing}

{While the proposed \ac{MQP} algorithm admits a closed-form solution for $\mathbf{x}$, it still involves a matrix inversion per outer iteration,
which can be computationally expensive for large systems, especially when the number of antennas is on the order of hundreds, as is the case in massive MIMO systems.}
In this section, we therefore introduce an \ac{MP}-based variation of the method, whose per-iteration cost scales as $\mathcal{O}(N_T N_U)$, i.e., linear on the number of \ac{BS} antennas $N_T$ and on the number of users $N_U$, with $N_U \ll N_T$ in massive MIMO, and which involves no matrix inversion. This is a significant reduction with respect to the closed-form update of Section~\ref{sec:proposed}, whose cost is
$\mathcal{O}(N_T^3)$ per outer iteration with standard matrix inversion, and at best
$\mathcal{O}(N_U^2 N_T)$ per outer iteration when the low-rank structure of $\mathbf{H}^H\mathbf{H}$ is exploited via the Woodbury identity\footnote{We quote the cost of standard matrix inversion, via Gaussian elimination or Cholesky factorization. Asymptotically faster algorithms exist---e.g., Strassen's algorithm~\cite{strassen1969gaussian} lowers the exponent to $\log_2 7 \approx 2.807$}.

As described in \cite{shrestha2026}, the discrete linear inversion problem in \cite{Iimori2021Robust} can be solved using a regularized \ac{GaBP} framework.
\ac{GaBP} is a general algorithm for performing Bayesian inference on a factor graph. {For the quantized precoding problem, we construct a dense factor graph representing the system,}
\begin{equation}
\mathbf{s} = \mathbf{H}\mathbf{x} + \mathbf{n}.
\label{eq:linear_model}
\end{equation}

{Let us emphasize that the model in \eqref{eq:linear_model} should not be confused with the physical receive model in~\eqref{eq:rx}, because in spite of both sharing the same linear form, they play dual roles.}
To elaborate, \ac{GaBP} works by iteratively passing messages along the edges of  this graph between \emph{variable nodes} (representing the unknown quantized signals $x_j$) and \emph{factor nodes} (representing the known symbols $s_i$).

{For this model, and to achieve a probabilistically consistent formulation, we normalize the entire objective by the noise variance to obtain}
\begin{align}
\operatorname*{minimize}_{\mathbf{x} \in \mathcal{C}_b^{N_T}}
\underbrace{\frac{P}{N_U\sigma_n^2}\|\mathbf{s}-\mathbf{H}\mathbf{x}\|_2^2}_{\propto -\log \mathbb{P}(\mathbf{s}|\mathbf{x})}
\!+\!
\underbrace{\|\mathbf{x}\|^2+\!\check{\lambda}\sum_{i=1}^{2^{2b}}\sum_{j=1}^{N_T}\gamma^2_{i,j}|x_j-c_i|^2.}_{\propto -\log \mathbb{P}(\mathbf{x})}
\label{eq:normalized_robust_cost_func}
\end{align}

{In the \ac{GaBP} framework, we approximate all messages as Gaussian distributions, parameterized by their mean and variance: $\mu \sim \mathcal{CN}(\text{mean}, \text{variance})$.} The algorithm proceeds in two main steps at each iteration:

\begin{enumerate}
\item \emph{Factor-to-Variable Message} ($\mu_{f_n \to v_m}$): This message represents the belief about $x_m$ from the $n$-th observation $s_n$, after marginalizing over all other variables $\{x_i \mid i \neq m\}$.
{Under the scalar Gaussian approximation, we approximate the interference term $I_{n,m} = \sum_{i \neq m} h_{n,i} x_i$ as a Gaussian variable. This leads to the \textit{soft interference-cancellation} (soft-IC) step, in which we compute the extrinsic mean as}
\begin{align}
\tilde{s}_{n,m}^{(t)} &= s_n - \sum_{i \neq m} h_{n,i} \hat{x}_{n,i}^{(t-1)}, \label{eqn:ext_mean} \\
\psi_{n,m}^{(t)} &= \sum_{i \neq m} |h_{n,i}|^2 \hat{v}_{n,i}^{(t-1)} + \sigma_n^2, \label{eqn:ext_var}
\end{align}
where $\tilde{s}_{n,m}^{(t)}$ is the soft interference-cancelled observation and $\psi_{n,m}^{(t)}$ is the corresponding conditional variance.

\item \emph{Variable-to-Factor Message} ($\mu_{v_m \to f_n}$): This message is the product of the prior $\mathbb{P}(x_m)$ and all incoming messages from factors other than $n$.
{First, we aggregate the incoming likelihood messages $\prod_{k \neq n} \mu_{f_k \to v_m}^{(t)}(x_m)$. This product is a complex Gaussian, $\mathcal{CN}(x_m; \bar{y}_{n,m}, \bar{v}_{n,m})$, with parameters}
\begin{eqnarray}
&\bar{v}_{n,m}^{(t)} = \left( \sum_{k \neq n} \dfrac{|h_{k,m}|^2}{\psi_{k,m}^{(t)}} \right)^{\!\!-1}\!\!\!\!,& \label{eqn:agg_var}\\
&\bar{y}_{n,m}^{(t)} = \bar{v}_{n,m}^{(t)} \left( \sum_{k \neq n} \dfrac{(h_{k,m})^{H} \tilde{s}_{k,m}^{(t)}}{\psi_{k,m}^{(t)}} \right).&
\label{eqn:agg_mean}
\end{eqnarray}
\end{enumerate}

\begin{figure*}[tp]
\centering
\includegraphics[width=1.4\columnwidth]{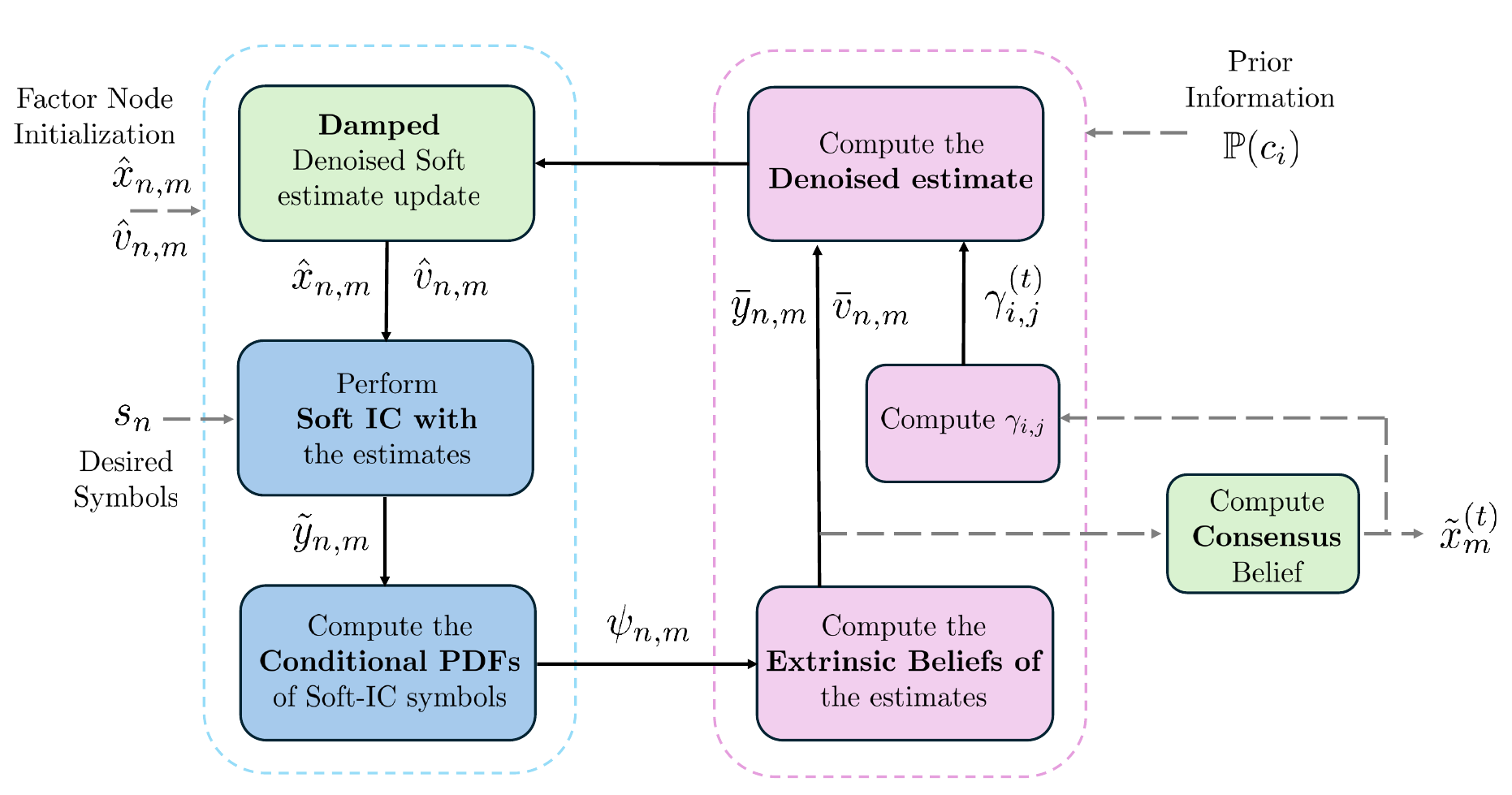}
\caption{Graphical representation of the Proposed GaBP-Quantized Precoding algorithm.}
\label{fig:flowchart}
\vspace{-2ex}
\end{figure*}

{While the marginalization suggests quadratic complexity per variable, it can be implemented efficiently using broadcast \cite{shental2008}, as shown later.}

{Crucially, we perform the denoising step by combining the aggregate belief with the normalized \ac{IDLS} prior.} The variable-to-factor update thus becomes a scalar minimization problem derived from \eqref{eq:normalized_robust_cost_func}
\begin{equation}
f(x_m) = \frac{1}{\bar{v}_{n,m}}|x_m - \bar{y}_{n,m}|^2 + |x_m|^2 + \check{\lambda} \sum_i \gamma_{i,m}^2 |x_m - c_i|^2.
\end{equation}

{To find the mean of the outgoing message, we apply the Wirtinger derivative, analogously to \ac{MQP},}
\begin{equation}
\frac{1}{\bar{v}_{n,m}}(x_m - \bar{y}_{n,m}) + x_m + \check{\lambda} \sum_{i} \gamma_{i,m}^2 (x_m - c_i) = 0.
\end{equation}

Solving for $x_m$ gives the denoising function yielding the updated message mean
\begin{equation}
\hat{x}_{n,m}^{(t)} = \frac{ \bar{y}_{n,m}^{(t)}/\bar{v}_{n,m}^{(t)} + \check{\lambda} \sum_{i} (\gamma_{i,m}^{(t)})^2 c_i }{ 1/\bar{v}_{n,m}^{(t)} + 1 + \check{\lambda} \sum_{i} (\gamma_{i,m}^{(t)})^2 },
\label{eqn:gabp_denoiser}
\end{equation}
while the corresponding message variance is
\begin{equation}
\hat{v}_{n,m}^{(t)} = \left( \frac{1}{\bar{v}_{n,m}^{(t)}} + 1 + \check{\lambda} \sum_{i} (\gamma_{i,m}^{(t)})^2 \right)^{\!\!-1}\!\!\!\!.
\label{eqn:gabp_var_upd}
\end{equation}

Unlike the closed-form solution, whose weight is selected by the discrepancy principle (Section~\ref{sec:lambda}), the \ac{GaBP} denoiser~\eqref{eqn:gabp_denoiser} needs no fine tuning of $\check{\lambda}$.
Instead, it suffices to make this parameter large enough for the discreteness prior to take effect.
With
\begin{equation}
G_m^{(t)} \triangleq \sum_i \big(\gamma_{i,m}^{(t)}\big)^2,
\qquad
\mu_m^{(t)} \triangleq
\frac{\sum_i \big(\gamma_{i,m}^{(t)}\big)^2 c_i}
     {\sum_i \big(\gamma_{i,m}^{(t)}\big)^2},
\label{eqn:disc_message}
\end{equation}
the update \eqref{eqn:gabp_denoiser} is the variable-node fusion of the data belief $\mathcal{CN}(\bar{y}_{n,m},\bar{v}_{n,m})$ and the discreteness message $\mathcal{CN}(\mu_m^{(t)},1/(\check{\lambda}G_m^{(t)}))$,
\begin{equation}
\hat{x}_{n,m}^{(t)} =
\frac{\bar{v}_{n,m}^{-1}\,\bar{y}_{n,m}
  + \check{\lambda}\,G_m^{(t)}\,\mu_m^{(t)}}
 {\bar{v}_{n,m}^{-1} + 1 + \check{\lambda}\,G_m^{(t)}},
\label{eqn:fusion_form}
\end{equation}
so that the precisions add and the means combine by inverse-variance weighting
\cite{Kay1993,shental2008,WeissFreeman2001}.

{For this fusion to work, the discreteness prior must be at least as informative as the data belief; that is, the condition $\check{\lambda}G_m^{(t)}\gtrsim\bar{v}_{n,m}^{-1}$ must hold. Since $G_m^{(t)}$ is small (the $\gamma_{i,m}$ are damped by $\zeta$), this calls for a sufficiently large $\check{\lambda}$.}
Once the prior is engaged, the \ac{BER} is essentially insensitive to the exact
value, {so we use a single large $\check{\lambda}$ throughout.}

\begin{algorithm}[H]
\caption{Proposed GaBP-Quantized Precoding}\label{alg:alg3}
\begin{algorithmic}
\State \textbf{External Input:}
\State \quad Symbols $\bm{s}$, channel matrix $\bm{H}$, noise power $\sigma_{\bm{n}}^2$, $\alpha$
\State \textbf{Internal Parameters:}
\State \quad Maximum iterations $k_{\max}$, damping factor $\rho \in (0, 1]$
\State \quad Convergence threshold $\varepsilon \ll 1$,
\State \textbf{Initialization:}
\State \quad Set $k = 0$, $\hat{\bm{x}}^{(0)} = \bm{0}_{N_T \times N_U}$,\; $\hat{\bm{v}}^{(0)} = \bm{1}_{N_T \times N_U}$
\State \textbf{repeat}
\State \quad Compute extrinsic mean $\tilde{\bm{s}}^{(k)}$ \hfill $\triangleright$ \eqref{eqn:ext_mean}
\State \quad Compute conditional variance $\bm{\Psi}^{(k)}$ \hfill $\triangleright$ \eqref{eqn:ext_var}
\State \quad Update $\gamma_{i,m}^{(k)}\ \forall\, i, m$ \hfill $\triangleright$ \eqref{eq:gamma}
\State \quad Compute aggregate variance $\bar{\bm{v}}^{(k)}$ \hfill $\triangleright$ \eqref{eqn:agg_var}
\State \quad Compute aggregate mean $\bar{\bm{y}}^{(k)}$ \hfill $\triangleright$ \eqref{eqn:agg_mean}
\State \quad Update message mean $\hat{\bm{x}}^{(k)}$ \hfill $\triangleright$ \eqref{eqn:gabp_denoiser}
\State \quad Update message variance $\hat{\bm{v}}^{(k)}$ \hfill $\triangleright$ \eqref{eqn:gabp_var_upd}
\State \quad Apply damping to $\hat{\bm{x}}^{(k+1)}$ and $\hat{\bm{v}}^{(k+1)}$ \hfill $\triangleright$ \eqref{eqn:gabp_damped_estimate}, \eqref{eqn:gabp_damped_var}
\State \quad Compute $\tilde{v}_m^{(k)}$ and $\tilde{x}_m^{(k)}\ \forall\, m$ \hfill $\triangleright$ \eqref{eqn:gabp_hard_var}, \eqref{eqn:gabp_hard_estimate}
\State \quad Compute $\varepsilon_k = \|\hat{\bm{x}}^{(k+1)} - \hat{\bm{x}}^{(k)}\|_2$
\State \textbf{until} $k > k_{\max}$ \textbf{or} $\varepsilon_k < \varepsilon$
\end{algorithmic}
\end{algorithm}
\vspace{-1ex}

{Finally, we apply a damping factor $\rho \in (0,1]$ to mitigate the oscillations common in loopy belief propagation, yielding}
\begin{align}
\hat{x}_{n,m}^{(t+1)} &= \rho\, \hat{x}_{n,m}^{(t)} + (1-\rho)\, \hat{x}_{n,m}^{(t-1)},
\label{eqn:gabp_damped_estimate} \\
\hat{v}_{n,m}^{(t+1)} &= \rho\, \hat{v}_{n,m}^{(t)} + (1-\rho)\, \hat{v}_{n,m}^{(t-1)}.
\label{eqn:gabp_damped_var}
\end{align}

{After each iteration, we perform a belief consensus to obtain a hard estimate $\tilde{x}_m^{(t)}$, which updates $\gamma_{i,m}$ for the next iteration,}
\begin{eqnarray}
&\tilde{v}_{m}^{(t)} = \Big( \sum_{k} \tfrac{|h_{k,m}|^2}{\psi_{k,m}^{(t)}} \Big)^{\!\!-1}\!\!\!\!,& \label{eqn:gabp_hard_var} \\
&\tilde{x}_{m}^{(t)} = \tilde{v}_{m}^{(t)} \Big( \sum_{k} \tfrac{(h_{k,m})^{H} \tilde{s}_{k,m}^{(t)}}{\psi_{k,m}^{(t)}} \Big).& \label{eqn:gabp_hard_estimate}
\end{eqnarray}

{Algorithm~\ref{alg:alg3} summarizes the resulting method.}

\vspace{-1ex}
\section{Simulation Results}
\label{sec:Simulations}

{We now present simulation results evaluating the performance of the proposed algorithms, in comparison with \ac{SotA} methods such as the quantized linear and nonlinear precoders based on \eqref{eq:jacobsson}.}

\begin{figure}[H]
\begin{center}
\subfigure[Small system with 1-bit quantization, with $N_T = 8$, $N_U = 2$, QPSK]{\includegraphics[width=\columnwidth]{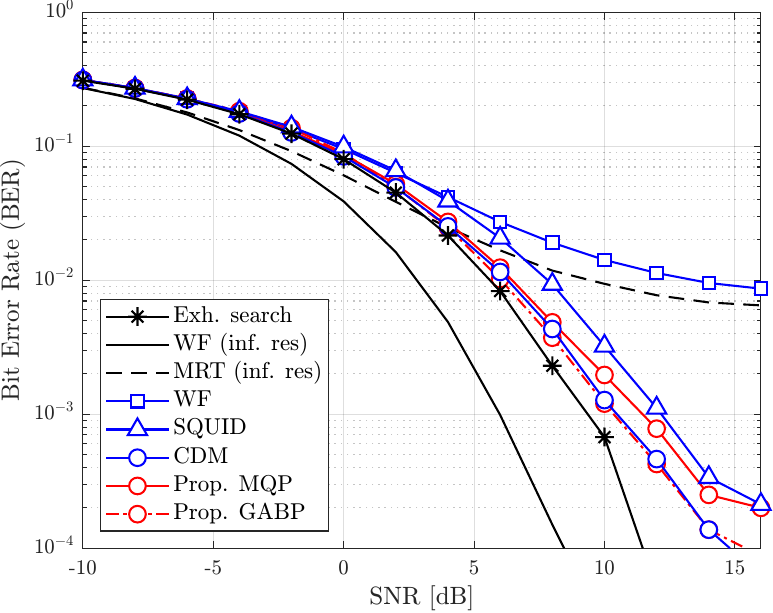}
\label{fig:ber_small_1bit}}
\subfigure[Small system with 2-bit quantization, with $N_T = 6$, $N_U = 2$ 16-QAM]{
\includegraphics[width=\columnwidth]{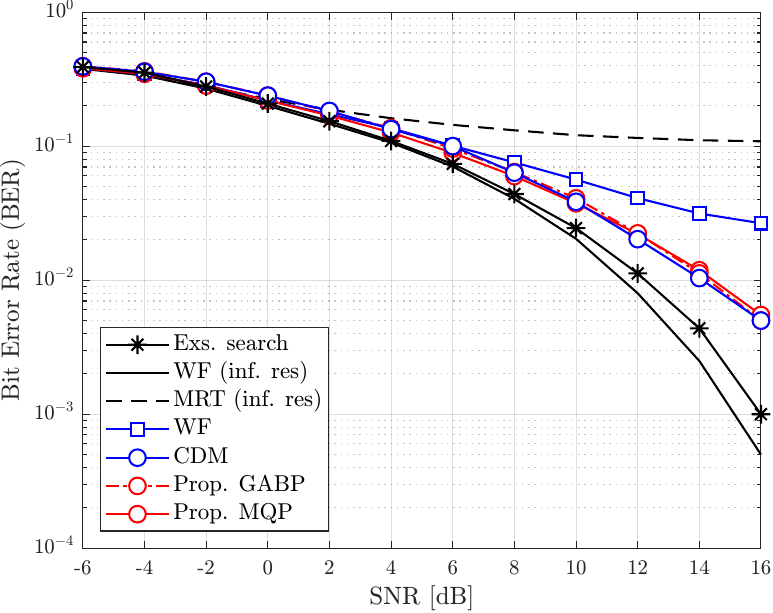}
\label{fig:ber_small_2bit}}
\end{center}
\vspace{-1.25ex}
\caption{BER performance of the proposed algorithms compared to the {SotA} and linear precoders, for small system sizes, where Exhaustive Search is feasible.}
\label{fig:ber_small}
\vspace{-4ex}
\end{figure}

{To that end, we consider a massive MIMO system with different numbers of transmit antennas $N_T$ and users $N_U$, and evaluate the \ac{BER} performance of the proposed algorithms under different \ac{SNR} levels and modulation schemes, such as QPSK and 16-QAM.}

For the small configurations in Fig.~\ref{fig:ber_small_1bit} (1-bit) and Fig.~\ref{fig:ber_small_2bit} (2-bit), the array is small enough for exhaustive search to be evaluated, which provides the optimal finite-alphabet benchmark; the proposed \ac{MQP} and \ac{GaBP} are near-optimal, remaining within a small gap of {exhaustive search}, whereas the quantized linear precoders (WF, MRT) exhibit an error floor, since they enforce the alphabet only by post-quantization rounding.
{As for the nonlinear baselines, CDM is competitive with the proposed algorithms, but both proposed algorithms outperform SQUID, since it does not account for the discrete nature of the transmit alphabet.}
Moreover, SQUID is not applicable to multi-bit alphabets, whereas the proposed algorithms extend naturally to any $b$-bit alphabet without modification.

\begin{figure}[H]
\centering
\subfigure[Large system with 1-bit quantization, with $N_T = 64$, $N_U = 8$, QPSK]{%
  \includegraphics[width=\columnwidth]{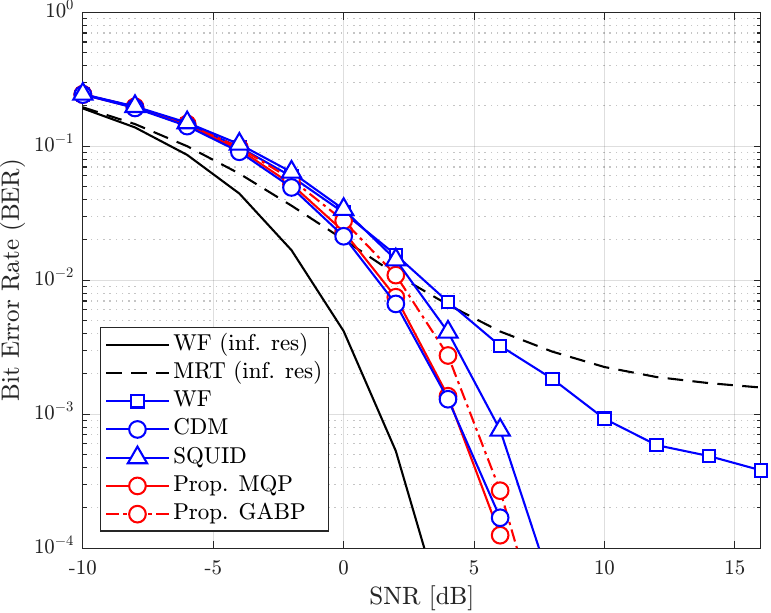}%
  \label{fig:ber_large_1bit}}\\
\subfigure[Large system with 2-bit quantization, with $N_T = 64$, $N_U = 8$, 16-QAM]{%
  \includegraphics[width=\columnwidth]{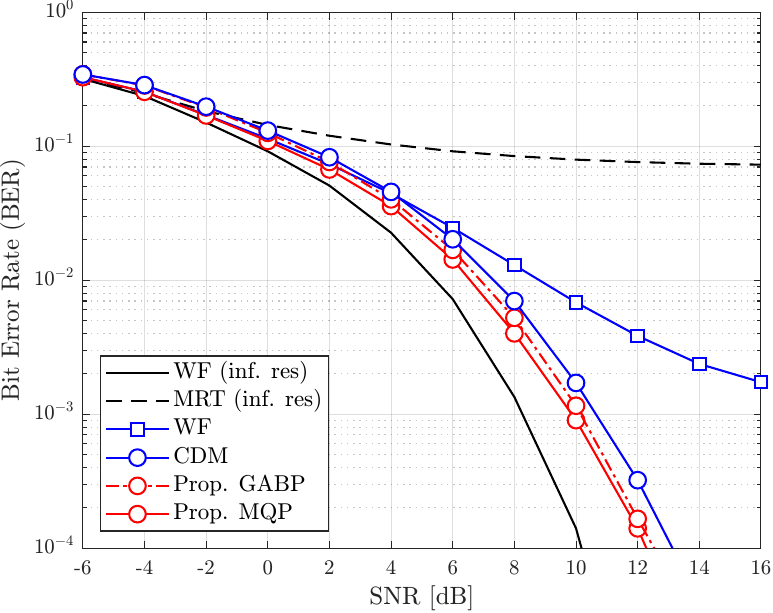}%
  \label{fig:ber_large_2bit}}
\caption{BER performance of the proposed algorithms compared to the {SotA} and linear precoders, for different levels of quantization and large system size. }
\label{fig:ber_large}
\end{figure}

For the larger $N_T=64$, $N_U=8$ arrays in Fig.~\ref{fig:ber_large_1bit} (1-bit) and Fig.~\ref{fig:ber_large_2bit} (2-bit), Exhaustive Search is no longer shown, as its complexity grows as $2^{bN_T}$ and cannot be evaluated at this scale.
Among the feasible schemes the proposed precoders stay closest to the infinite-resolution bound and outperform CDM and SQUID, confirming that their advantage persists precisely in the regime where the optimum is out of reach.

\begin{figure*}
\centering
\includegraphics[width=0.99\textwidth]{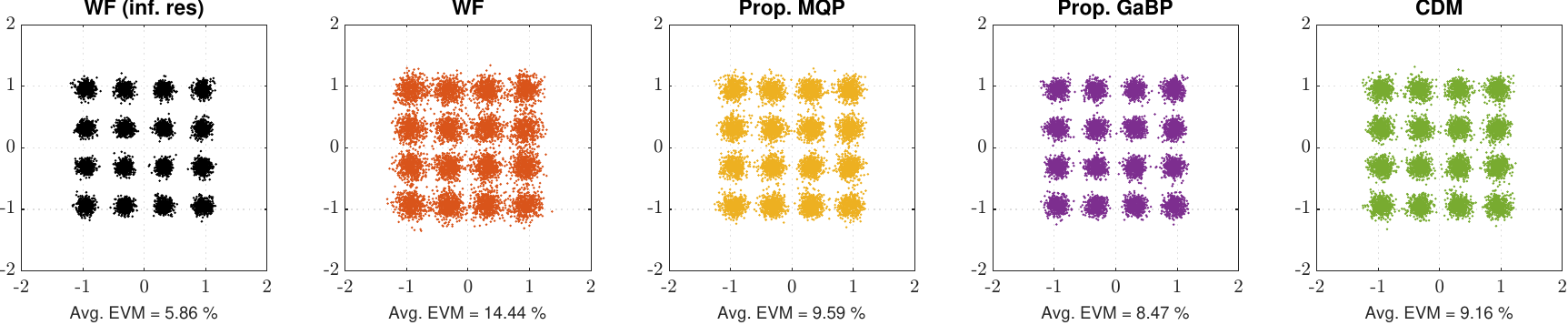}
\caption{Constellation diagram of the proposed MQP algorithm for a massive MIMO system with $N_T = 64$ and $N_U = 8$, ${P_{\text{avg}} = 16 \text{ dB}}$ under 16QAM modulation, with a {2-bit} quantizer.
}
\label{fig:ber_constellation}
\end{figure*}

Increasing the DAC resolution from one to two bits narrows the gap to infinite resolution for all schemes, but the proposed precoders extend to multi-bit alphabets without modification, whereas the linear baselines require additional bits to attain comparable reliability.
These trends are confirmed in Fig.~\ref{fig:ber_constellation}, where the proposed precoders produce tighter symbol clusters and a lower \ac{EVM} than linear methods, reflecting their higher fidelity at the same transmit power.

\subsubsection{Channel estimation error}

{Next, we evaluate the proposed algorithms under channel estimation errors, a common impairment in practical systems that can significantly degrade precoding performance if not properly accounted for, comparing them against the quantized linear and nonlinear precoders.}
{To this end, we model the channel estimation error as a Gauss-Markov noise matrix~$\mathbf{E}$ added to the true channel matrix $\mathbf{H}$, yielding the estimated channel matrix}
\begin{equation}
\hat{\mathbf{H}} = \sqrt{1-\tau^2}\mathbf{H} + \tau \mathbf{E},
\end{equation}
{where $\tau\in [0,1]$ controls the level of channel estimation error.}

As shown in Fig.~\ref{fig:ber_csi}, the proposed precoders degrade gracefully as the relative CSI error $\tau$ grows and retain their margin over the baselines, indicating that the discrete-aware design is robust to imperfect channel knowledge.
{Notably, the proposed algorithms also outperform the \ac{CDM} precoder, which is regarded as the best-performing nonlinear precoder in the literature and is itself robust to channel impairments.}
{Furthermore, the proposed algorithms, particularly \ac{MQP}, can compensate for channel estimation errors by modifying the regularization term in \eqref{eq:l0} to account for the channel uncertainty, which offers a route to further improving performance under imperfect \ac{CSI}.}

\subsubsection{Spatial channel correlation}

{Finally, we assess robustness to spatially correlated channels using the Kronecker
model}
\begin{equation}
\mathbf{H} = \mathbf{R}_{\mathrm{rx}}^{1/2}\,\mathbf{H}_{\mathrm{w}}\,\mathbf{R}_{\mathrm{tx}}^{1/2},
\label{eq:kron}
\end{equation}
where $\mathbf{H}_{\mathrm{w}}$ has i.i.d.\ $\mathcal{CN}(0,1)$ entries and $\mathbf{R}_{\mathrm{tx}}$, $\mathbf{R}_{\mathrm{rx}}$ are the transmit- and receive-side correlation matrices. {Since the users are single-antenna terminals, we set $\mathbf{R}_{\mathrm{rx}}=\mathbf{I}$ and model the correlation across the \ac{BS} uniform linear array by the classical Jakes/Clarke spatial correlation under isotropic scattering \cite{Jakes1974},}
\begin{equation}
[\mathbf{R}_{\mathrm{tx}}]_{p,q} = J_0\Big(2\pi \frac{d}{\lambda}\,|p-q|\Big),
\label{eq:jakes}
\end{equation}
where $J_0(\cdot)$ is the zeroth-order Bessel function of the first kind and $d/\lambda$ is the antenna spacing in wavelengths.

\begin{figure}[H]
\centering
\subfigure[Channel estimation impairment, for $N_T = 64$, $N_U = 8$, 16QAM]{%
  \includegraphics[width=\columnwidth]{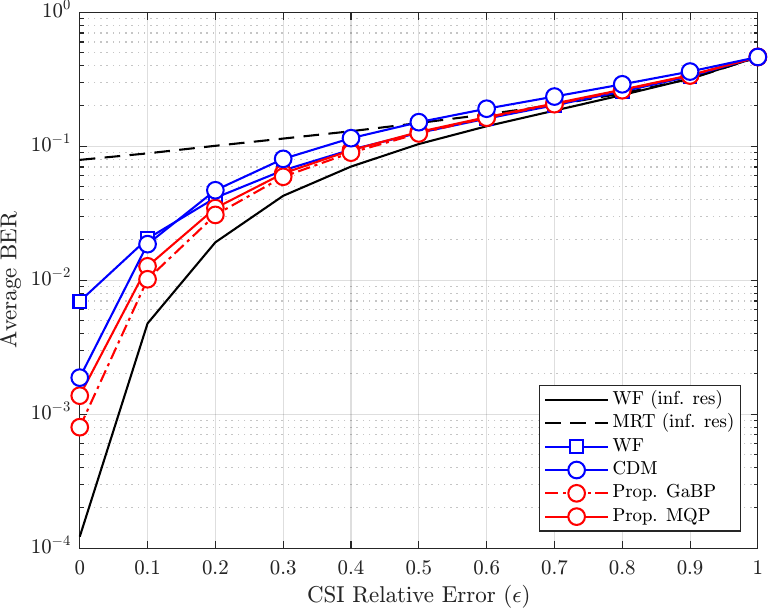}%
  \label{fig:ber_csi}}\\
\subfigure[Spatial correlation impairment, for $N_T = 16$, $N_U = 4$, 16QAM]{%
  \includegraphics[width=\columnwidth]{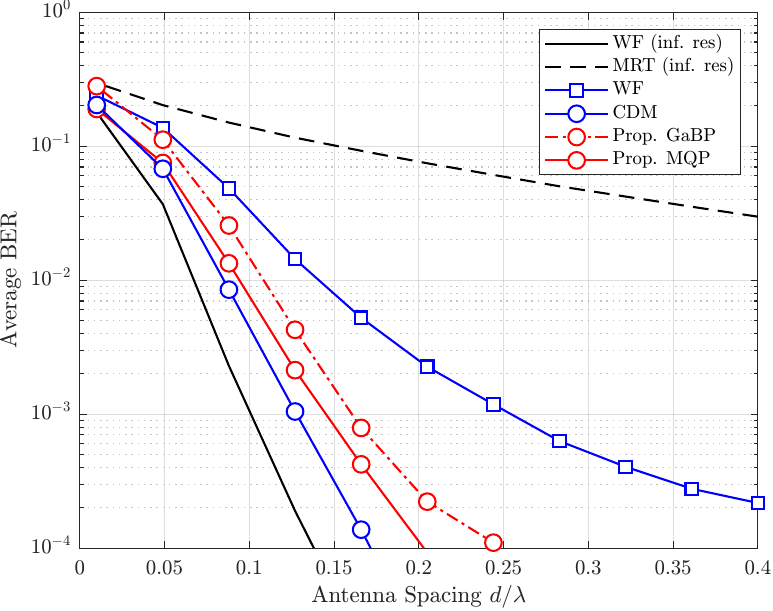}%
  \label{fig:ber_corr}}
\caption{BER performance of the proposed algorithms compared to the {SotA} and linear precoders, under a) channel estimation errors and b) spatially correlated channels.}
\label{fig:ber_robust}
\vspace{-1ex}
\end{figure}

Notice that the correlation increases as the spacing shrinks. Namely, nearest-neighbor correlation is $\rho = J_0(2\pi d/\lambda)$, so that $d/\lambda \approx 0.4$ is weakly correlated while $d/\lambda \to 0$ is strongly correlated.
Consider therefore Fig.~\ref{fig:ber_corr}, which shows the average \ac{BER} as a function of the antenna spacing at a fixed transmit \ac{SNR} of $15$\,dB, for $N_T=16$, $N_U=4$, 16QAM signaling, and $2$-bit DACs.
As the spacing decreases and correlation grows, the error rate of all precoders rises; the proposed \ac{GaBP} precoder degrades gracefully and tracks the closed-form \ac{MQP}, retaining its advantage over the linear baselines across the entire correlation range.
However, both algorithms are sensitive to strong correlation, which is expected since the channel matrix becomes ill-conditioned and the effective rank drops, making the finite-alphabet precoding problem more difficult, while \ac{CDM} is more robust to correlation, as it greedily updates one antenna at a time, exploiting the residual to mitigate the correlation effects.

{Nonetheless, for the first algorithm, we can compensate for the spatial correlation by modifying the regularization term in \eqref{eq:l0} to account for the correlation level, which we can estimate from the channel matrix $\mathbf{H}$; this offers a route to further improving performance under spatially correlated channels.}
As for the second algorithm, there are several techniques to mitigate the effects of spatial correlation, such as \cite{Takahashi2019ASB, Shirase2020} which can be incorporated into the proposed \ac{GaBP} precoder to further enhance its performance under correlated channels.
\vspace{-1ex}
\subsection{Complexity Analysis}

{Finally, we compare the per-iteration cost of the two proposed solvers.}
{In the closed-form \ac{MQP}, each iteration solves the regularized linear system
\eqref{eq:closed_mmse}, whose system matrix is $\mathbf{B}+\check\lambda\mathbf{G}$ with
$\mathbf{B}=\mathbf{H}^{H}\mathbf{H}+\tfrac{N_U\sigma_n^2}{P}\mathbf{I}$; a direct inversion
would cost $\mathcal{O}(N_T^3)$.
Crucially, the diagonal penalty $\check\lambda\mathbf{G}$ is full-rank and changes at every
iteration, so no factorization of $\mathbf{B}$ can be reused. Instead, we exploit the low
rank of the Gram term: $\mathbf{H}^{H}\mathbf{H}$ has rank at most $N_U$, while the remaining
part $\mathbf{A}\triangleq\tfrac{N_U\sigma_n^2}{P}\mathbf{I}+\check\lambda\mathbf{G}$ is diagonal
and hence inverted in $\mathcal{O}(N_T)$. Applying the matrix-inversion lemma to the low-rank
term $\mathbf{H}^{H}\mathbf{H}$ recasts each solve as the inversion of an $N_U\times N_U$
matrix, reducing the per-iteration cost to $\mathcal{O}(N_U^{2}N_T)$.}

\begin{figure}[H]
\centering
\includegraphics[width=\columnwidth]{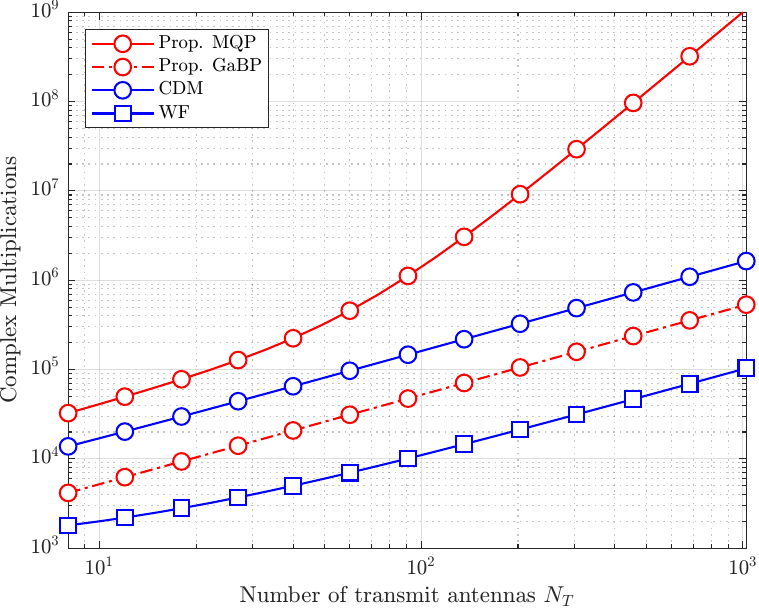}
\caption{Number of complex multiplications versus number of transmit
antennas $N_T$ for the proposed and \ac{SotA} algorithms, for $N_U = 10$,
16-QAM, a 2-bit quantizer, at SNR $=10$\,dB\protect\footnotemark.}
\label{fig:complexity}
\end{figure}

The auxiliary \ac{FP} weights $\gamma_{i,j}$ add $\mathcal{O}(N_T\,2^{2b})$, and the discrepancy update of $\lambda$ reuses the residual already formed, costing only $\mathcal{O}(1)$ scalar operations.
The message-passing formulation removes the matrix inversion altogether.
{Each \ac{GaBP} iteration consists solely of element-wise products and row/column reductions over the $N_U\times N_T$ factor graph---the soft interference cancellation \eqref{eqn:ext_mean}, \eqref{eqn:ext_var}, the aggregation \eqref{eqn:agg_var}--\eqref{eqn:agg_mean}, and the denoiser \eqref{eqn:gabp_denoiser}---together with the $\mathcal{O}(N_T\,2^{2b})$ discreteness weights.}
Its per-iteration cost is therefore $\mathcal{O}(N_T N_U)$, {i.e.,}\ linear in the number of antennas.
Over $T$ iterations the two solvers cost {$\mathcal{O}(N_T^3 + T N_U^2 N_T)$} and $\mathcal{O}(T N_T N_U)$, respectively, so the gap widens as the array grows.
This makes the \ac{GaBP} precoder especially attractive for massive MIMO, where $N_T$ is large, while, as the preceding simulations show, it attains a \ac{BER} close to the closed-form solution and the \ac{SotA} nonlinear precoder.
\Ac{CDM} is inversion-free in its main loop: after an MMSE-type initialization (a one-time $N_U\times N_U$ inverse) it greedily updates one antenna at a time against an incrementally maintained residual, at $\mathcal{O}(N_T N_U 2^{2b})$ per sweep.
\begin{table}[tp]
\centering
\caption{Asymptotic complexity of $b$-bit quantized precoders.}
\label{tab:complexity}
\renewcommand{\arraystretch}{1.3}
\begin{tabular}{@{}lc@{}}
\toprule
\textbf{Precoder} & \textbf{Complexity} \\
\midrule
Linear (WF)        & $\mathcal{O}\!\left(N_U^{3} + N_U^{2} N_T\right)$ \\
SQUID  \cite{Jacobsson2017quantized}                    & $\mathcal{O}\!\left(N_U^{3} + T\,N_T N_U\right)$ \\
CDM                 & $\mathcal{O}\!\left(N_U^{3} + T\,2^{2b} N_T N_U\right)$\\
{Proposed MQP}       & $\mathcal{O}\!\left(N_U^{3} + T\,N_U^{2} N_T\right)$ \\
Proposed GaBP                & $\mathcal{O}\!\left(T\,N_T\,(N_U + 2^{2b})\right)$ \\
Exhaustive search         & $\mathcal{O}\!\left(N_U\,2^{\,2b N_T}\right)$ \\
\bottomrule
\end{tabular}
\vspace{-2ex}
\end{table}

Both baselines are thus linear in $N_T$ like the proposed \ac{GaBP} precoder, but, unlike the latter, still rely on a precomputed matrix inverse.

\footnotetext{{We choose the parameter $T$ in Table~\ref{tab:complexity} for each iterative algorithm based on the convergence figure below, ensuring a fair comparison across all methods.}}

{Table~\ref{tab:complexity} summarizes the computational complexity orders of all considered algorithms, both \ac{SotA} and proposed.}
For a graphical view, Figure~\ref{fig:complexity} plots the asymptotic
complexity expressions (in Complex Multiplications) as a function of $N_T$, with constant factors omitted. Therefore, the figure illustrates the scaling behavior of each algorithm rather than a literal count of operations.

\subsection{Convergence Analysis}
\label{sec:convergence}
{This section examines the convergence and nature of the proposed and \ac{SotA} solutions.}
As the finite-alphabet problem~\eqref{eq:l0} is combinatorial, neither solver can guarantee the global optimum; the relevant guarantees concern convergence to a stationary point of the smooth surrogate.

\subsubsection{Proposed \ac{MQP}} For fixed regularization weight $\lambda$ and smoothing
parameter $\zeta$, the \ac{MQP} iteration is a \ac{MM} scheme: the auxiliary update $\gamma_{i,j}=\sqrt{\zeta}/(|x_j-c_i|^2+\zeta)$ is the exact maximizer of the quadratic transform of the smoothed $\ell_0$ surrogate, and the $\mathbf{x}$-step minimizes the resulting convex quadratic \eqref{eq:quadratic} in closed form \cite{Shen2018,Sun2017}.
Three facts establish convergence:
\emph{(i)~Well-posedness:} the per-iteration matrix in \eqref{eq:closed_mmse} is the sum of the positive-semidefinite $\mathbf{H}^{H}\mathbf{H}$, the strictly positive-definite $\tfrac{N_U\sigma_n^2}{P}\mathbf{I}$, and the positive-semidefinite diagonal $\check{\lambda}\mathbf{G}$; hence this matrix is positive definite \emph{irrespective} of the rank or conditioning of~$\mathbf{H}$, and the update has a unique solution at every iteration.
\emph{(ii)~Monotonicity:} each block update, i.e., $\gamma$, $\mathbf{x}$, and the optimal scaling $\beta^\star$, exactly minimizes its block, so the smoothed objective is non-increasing across iterations.
\emph{(iii)~Boundedness:}
the objective is bounded below by zero.
A monotone, bounded-below sequence converges, and by standard MM arguments \cite{Sun2017} every limit point of the iterates is a stationary point of the (non-convex) smooth objective.
Hence the objective cannot diverge or oscillate---though, as for any \ac{MM} scheme, this does not by itself guarantee a convergence rate or finite-time termination.

\begin{figure}[H]
\centering
\includegraphics[width=\columnwidth]{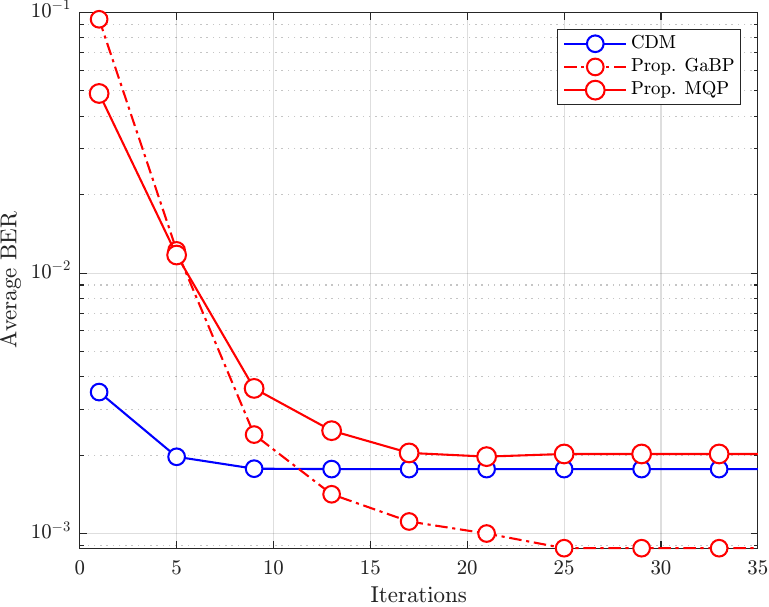}
\caption{Convergence of the proposed algorithms, for $N_T = 64$ and $N_U = 8$, under 16QAM modulation, with a {2-bit} quantizer, at SNR = 10 dB.}
\label{fig:convergence}
\end{figure}

The discrepancy update of $\lambda$ and the graduated-non-convexity annealing of $\zeta$ make the objective vary across stages, so this guarantee applies once they settle to $(\lambda^\star,\zeta_{\min})$; the algorithm returns a local optimum of \eqref{eq:l0}.

\subsubsection{Proposed \ac{GaBP}.} {The \ac{GaBP} solver is a loopy belief-propagation scheme with a nonlinear discreteness denoiser~\eqref{eqn:gabp_denoiser}, for which a \emph{universal} convergence guarantee is not known~\cite{Weiss2000}; its behavior is nonetheless well characterized.}
For the underlying linear-Gaussian model, \ac{GaBP} returns the exact posterior
means whenever it converges \cite{WeissFreeman2001}, and a sufficient condition for
convergence is walk-summability---equivalently, generalized diagonal
dominance---of the precision matrix \cite{Malioutov2006}.

In the message-passing regime the iteration is governed by state evolution and
converges for i.i.d.\ sub-Gaussian $\mathbf{H}$ in the large-system limit
\cite{Bayati2011}.
For finite or spatially correlated $\mathbf{H}$, however, message passing may diverge,
which can nevertheless be mitigated via damping -- namely, $\rho\in(0,1]$, as in
\eqref{eqn:gabp_damped_estimate}--\eqref{eqn:gabp_damped_var}, or, under strong
correlation, by scaled-belief variants.
{When convergent, the fixed point of the message-passing solver is a stationary point of the same regularized objective targeted by \ac{MQP}, so that the two solvers approximate the same local
solution---\ac{GaBP} at linear per-iteration cost.}
{Figure~\ref{fig:convergence} compares the convergence of the proposed methods with the CDM-based \ac{SotA} adaptation.}
In all reported experiments the damped iteration converged within the iteration
budget.

\section{Conclusions}
\label{sec:conclusion}
We have proposed two novel precoding algorithms for massive MIMO systems with low-resolution DACs, based on a discrete-aware formulation of the finite-alphabet precoding problem.
The first algorithm, \ac{MQP}, is a closed-form solution that leverages the quadratic transform to decouple the non-convex objective into a sequence of convex subproblems, while the second algorithm, \ac{GaBP}, is a message-passing scheme that approximatses the same objective at linear per-iteration cost.
Both algorithms outperform existing linear and nonlinear precoders in terms of \ac{BER} and \ac{EVM}, and are robust to channel estimation errors and spatial correlation.
The proposed algorithms are also scalable to large antenna arrays, making them potentially suitable for practical deployments of massive MIMO wireless systems.

% \bibliographystyle{IEEEtran}
% \bibliography{references}

\balance

\end{document}